# Temperature Accelerated Life Test and Failure Analysis on Upright Metamorphic $Ga_{0.37}In_{0.63}P/Ga_{0.83}In_{0.17}As$/Ge Triple Junction Solar Cells

Vincenzo Orlando[1], Iván Lombardero[1], Mercedes Gabás[2], Neftali Nuñez[1], Manuel Vázquez[1], Pilar Espinet-González[1], Jesús Bautista[1], Rocio Romero[2], Carlos Algora[1]

[1] *Instituto de Energía Solar, Universidad Politécnica de Madrid, Madrid, 28040, Spain*
[2] *Dept. Física Aplicada I, The Nanotech Unit, Universidad de Málaga, Málaga, 29071, Spain*

**ABSTRACT**

A temperature accelerated life test on Upright Metamorphic $Ga_{0.37}In_{0.63}P/Ga_{0.83}In_{0.17}As$/Ge triple-junction solar cells has been carried out. The acceleration has been accomplished by subjecting the solar cells to temperatures (125, 145 and 165°C) significantly higher than the nominal working temperature inside a concentrator (90°C), while the nominal photo-current (500×) has been emulated by injecting current in darkness. The failure distributions have been fitted to an Arrhenius–Weibull model resulting in an activation energy of 1.39 eV. Accordingly, a 72 years warranty time for those solar cells for a place like Tucson (AZ, USA), was determined. After the ALT, an intense characterization campaign has been carried out in order to determine the failure origin. We have detected that temperature soak alone is enough to degrade the cell performance by increasing the leakage currents, the series resistance, and the recombination currents. When solar cells were also forward biased an increase of series resistance together with a reduction of short circuit current is detected. The failure analysis shows that: a) several metallization sub-products concentrate in several regions of front metal grid where they poison the silver, resulting in a two times reduction of the metal sheet resistance; b) the metal/cap layer interface is greatly degraded and there is also a deterioration of the cap layer crystalline quality producing a huge increase of the specific front contact resistance, c) the decrease of short circuit current is mainly due to the GaInP top subcell degradation.

## 1. INTRODUCTION

The deployment of Concentration Photovoltaic (CPV) systems based on Multijunction Solar Cell (MJSC) is currently dominated by systems integrating cell-on-carriers (CoCs) based on lattice-matched (LM) GaInP/Ga(In)As/Ge triple-junction concentrator solar cells .Their long term reliability has been assessed by means of temperature accelerated life tests (ALT) and failure analysis [1,2,3] proving a lifetime long enough in order not to limit the system one [3]. This type of LM solar cells is currently close to its practical efficiency limit [1], so no additional efforts are being paid to improve its performance, excepting those related to a further cost reduction. Because of this, most manufacturers and research institutes are currently investigating novel solar cell architectures able to achieve higher efficiencies [1]. The determination of the reliability of these novel architectures is a key factor to push their development and increase their market share.

One of these novel architectures is the Upright Metamorphic (UMM). Several studies have been carried out on the suitability of solar cells based on this architecture [4,5] which nowadays is starting to be fabricated for both terrestrial and space applications. Few papers [4,6,7] have treated the performance evolution of UMM solar cells when subjected to some degradation factors but only one has presented the preliminary results of a temperature ALT designed to determine the reliability of UMM $Ga_{0.37}In_{0.63}P/Ga_{0.83}In_{0.17}As/Ge$ triple-junction solar cells [7]. Accordingly, the purpose of this work is to show both the complete statistical reliability analysis and the failure analysis for this kind of UMM triple-junction solar cells.

## 2. ACCELERATED LIFE TEST

The CoCs with the UMM cells (see Figure 1 top left and right) were manufactured by Fraunhofer ISE (Germany) and were designed to be mounted inside a concentrator module using Fresnel lenses at 500× where the nominal working temperature is 90°C [8]. The maximum temperature the CoC can withstand is 170°C (limited by the solder stability). The CoCs were then sent to the Solar Energy Institute of the Universidad Politécnica de Madrid (IES-UPM, Spain) to carry out the pre-test characterization and the ALT. Additionally some characterization for the failure analysis was also carried out by the University of Málaga (UMA, Spain).

The first step to carry out the ALT is to assess the performance of the devices before the degradation process. Forty-eight solar cells were carefully characterized by means of dark I-V curve, 1× and 500× illumination I-V curve, electroluminescence (EL) mapping, and external quantum efficiency (EQE). These measurements were used to quantify the degradation observed in the devices throughout the ALT and post-test analysis, and to support the failure analysis process. Figure 2 (top left) shows the illuminated I-V curves at 500× of the solar cells before the ALT. The dispersion for both short circuit current ($I_{sc}$) and open circuit voltage ($V_{oc}$) is less than 3% while the dispersion for the maximum power point ($P_{MPP}$) is less than 5%. Figure 2 (top right) shows the EQE, and Figure 2 (bottom) shows the dark I-V curve for the solar cells before the ALT. Notice that the dark I-V curve of some cells exhibit a low shunt resistance ($R_P$) at very low currents probably due to accidental processing differences among devices. This signature is far away from the operation point at 500× where the devices have to be tested for reliability purposes

The temperature ALT was carried out in three climatic chambers, with 16 CoCs in each chamber. Each CoC was previously fitted with a 70 mm x 70 mm x 3 mm aluminum heatsink to reduce the junction temperature and to increase the thermal inertia of the system (see Figure 1 bottom left and right). From now on we refer to the CoCs as solar cells. For the ALT the photogenerated current was emulated by forward biasing the solar cells. Simulations with a 3D distributed model [2,9] were carried out to emulate nominal photo-current conditions at 500×, while avoiding electrical overstress that would produce artificial failures [2], resulting in an injected current of 340 mA (the active area of the solar cells was 0.0425 cm$^2$). The suitability of using forward bias current to emulate the photogenerated current in accelerating aging tests is explained in [2,10]. Finally, the failure criterion was defined as an illumination power loss of 10% at 500×. This value was defined taking into account the typical 5% error in the efficiency measurement of a multijunction solar cell together with the fact that a typical 20% power loss could be defined for the whole CPV system which includes optics, tracking, etc. that could fail.

Thermal finite element analysis (FEA) simulations were carried out to obtain an approximate value of the solar cell temperature under forward bias inside the climatic chamber. After that, the d$V$/d$T$ of several solar cells per chamber was measured using a digital multimeter and a controlled current source, this value was 5.5 mV/C ± 0.5 mV/C. The average value of the calculated d$V$/d$T$ was used then to obtain the junction temperature of the solar cells inside the chamber. This method has been successfully applied in previous ALTs [2,3] and is further explained in [11]. With this information the temperature of the climatic chambers was properly set in order to get 125, 145 and 165 °C as junction temperatures of the solar cells ($T_{cell}$) with current injection.

In order to periodically monitor the degradation of the solar cells inside the climatic chamber during the ALT, they were disconnected from the current injection to measure their dark I-V curve once they have stabilized to the temperature of the chamber. Once all the solar cells were measured, the current injection started again (this is defined as a "cycle"). Six solar cells (two solar cells per chamber) were not forward biased (no current injection) and were used as "reference" solar cells. For three of them called "Ref_Low", "Ref_Mid", and "Ref_High" (see Table 1) their I–V curve is measured only at the beginning and the end of the ALT. The purpose of these cells was to analyse only the effects of temperature excluding both the influence of current biasing and I-V curve measurements. Other three solar cells, called "Ref_IV_Low", "Ref_IV_Mid" and "Ref_IV_High", were measured after each degradation cycle in order to track their evolution as consequence of temperature impact. The purpose of these cells is to continuously monitor the impact of $T$ with the addition of periodical I-V curve measurement.

The nomenclature used in Table 1 comes from changes observed in the dark I-V curves after the degradation tests. Figure 3 shows the typical evolution of a degraded solar cell (in this particular case, "Fail_High" of Table 1) where the two main degradation processes are: (1) an increase in the recombination current, and (2) an increase of series resistance. Figure 3 also shows a shunt resistance variation which improves/worsens between cycles but without affecting the performance of the solar cell at 500×. Therefore, the shunt resistance variation is not relevant for the reliability analysis considering the nominal working conditions. Since two degradation mechanisms are observed, it is important to understand the time evolution of the dark I-V curve to analyze which degradation mechanism is dominating the solar cell performance. Figure 4 shows the evolution of the dark voltage at a current equivalent to the $I_{mpp}$ for several "Fail_High" solar cells. The first degradation process that appears is the increase in recombination currents, followed afterwards by an increase in series resistance, as shown in Figure 4.

The dark I-V curves provide a lot of information on the degradation of the solar cells, but they do not allow by themselves to determine when the solar cell has failed (notice that we have stated above the solar cell failure as the illumination power loss greater than 10% at 500×). Therefore, in order to connect the dark I-V curve evolution with the illumination power loss some solar cells were taken out of the climatic chambers during the ALT to quantify their degradation (see Table 1). Figure 5 shows the typical illumination I-V curves at 500× for a degraded cell at 165ºC.

By combining dark and illumination I-V curves it is possible to correlate the power loss under illumination with the voltage variation in darkness. Figure 6 shows this correlation for several "Fail_High", "Fail_Mid" and "Deg_Mid_Rs" solar cells when a dark voltage variation of 1.2 ± 0.2% corresponds to a 10% illumination power loss (which is the failure criterion). This correlation was used to determine the moment on which a solar cell had failed. Accordingly, when a solar cell inside the climatic chamber had a dark voltage variation of 1.2% the total hours were recorded, and the solar cell was regarded as a failure. Additionally, the failed solar cell was disconnected from the current source and was taken out from the climatic chamber for characterization.

After 8,900 hours of test time (time for which the solar cells where under forward bias): a) all the solar cells on the 165 °C test had failed (the last failure at 165 °C happened after ~4,000 hours, see blue dots in Figure 7), b) three solar cells on the 145 °C test had failed also, and c) there were no failures on the 125 °C test. Using the statistical analysis (see Section 3) the probable time for the next failure on the 145 °C test was calculated, which resulted in around 1,500 additional hours of test time (approximately 3 additional months of testing, see Section 3), and the probable time for the first failure at 125 °C test resulted in around 40,000 hours (approximately 48 additional months of testing!) as shown in Figure 7. Continuing the test was impractical, so the test was stopped to start the reliability and failure analysis, whose results are shown in Section 3 and 4, respectively.

## 3. RELIABILITY ANALYSIS

Once the tests were finished and failure times for each solar cell were recorded following the selected failure criterion (10% power loss at 500×), they were adjusted to a Weibull failure distribution function. The Weibull distribution with two parameters ($\beta$ –shape parameter- and $\eta$ –scale parameter-) was chosen to describe the life of solar cells due to its versatility, and moreover, it has demonstrated very good fits in previous temperature ALTs carried on concentrator III-V solar cells [2,12]. The life-stress model used in this work is the Arrhenius model, which is widely used to predict the life of a device when the applied stress is temperature [2,12]. The details of these statistical models and parameters can be found in [1].

Figure 7 shows the Weibull failure probability plot for the three temperature ALTs together with the extrapolated results at 90 °C (nominal working temperature) with 95% confidence levels [13]. Confidence levels provide information about the certainty of a value. A 95% confidence level means that 95% of the samples are contained in this interval and consequently, 95% of the solar cells will have a reliability equal or better to the number shown. Depending on the application, there are wide possibilities of using different confidence levels, 95% being the most used in experimental research, due to its conservative nature [14,15].

The failure time data in the Weibull plot of Figure 7 shows that the model fitted by the maximum likelihood estimation (MLE) method reproduces the experimental data of the ALT with a very good fit.

The values of the parameters obtained after fitting the Arrhenius-Weibull model [2] are:

- Weibull shape parameter, $\beta$: 2.39
- Arrhenius activation energy, $E_A$: 1.39 eV
- Weibull scale parameter, $\eta$, at 90 ºC, 95% confidence level: $6.4 \cdot 10^5$ h

A Weibull shape parameter larger than one (as in this case) implies that solar cell has a failure rate increasing with time which corresponds with degradation-like failures (wear out period of life) [1]. An Arrhenius-Weibull activation energy of 1.39 eV is meaningful, since the activation energies in III–V solar cells and optoelectronic devices range from 0.5 to 1.75 eV [16-18], and previous ALT results for concentrator III-V multijunction solar cells were 1.58 and 1.02 eV [2,12].

It is possible to evaluate warranty times from the reliability data. Warranty times are selected considering the percentage of failures allowed during warranty time period. If a manufacturer has a warranty policy that allows a 5% of failures during the warranty time period, this period will coincide with the time at which a reliability of 0.95 is achieved. Using the results of Figure 7 and the procedure explained in [2] it is possible to extrapolate the ALT results to nominal temperature (90 °C), using a very conservative confidence level (95%), obtaining a warranty time for a failure population of 5%, W$t$(5% failures) = 228,931 hours.

Extrapolating these warranty times to real working time requires knowing the location where the CPV system incorporating these Solar cells will operate, together with the subsequent information about climatic conditions. Applying the model of [19] with the specifications of these Solar cells, using the Direct Normal Irradiance (DNI), ambient temperature and wind speed for Tucson (AZ-USA), which is an example of good location for CPV, together with the reliability values calculated in this paper, a warranty time of 72 years for these UMM Solar cells is achieved. Other failures modes that are not detected with the current temperature ALT may appear under different weather and/or working conditions, such as moisture or thermal cycling [1], thus reducing the warranty time forecast.

4. FAILURE ANALYSIS

Once the solar cells failed and the reliability analysis was accomplished, failure analysis was carried out in order to determine the physical origins behind their degradation.

Several solar cells were selected for an extensive material characterization before and after the ALT, namely scanning electron microscopy (SEM), transmission electron microscopy (TEM) and energy-dispersive X-ray spectroscopy (EDX). More specifically, the solar cell front metal grid evolution was monitored using SEM. Planar view and cross-section images of several degraded and non-degraded solar cells were taken with a FEI Helios Nanolab 650 Dual Beam microscope, fitted with a Schottky field emission source for SEM (FESEM) and a Tomahawk focused ion beam (FIB), which allows a precise and reliable milling and patterning. EDX was performed in several planar view and cross-section cell areas, including the metallization grid and the busbars, to monitor their chemical composition. Different areas of the cells were analyzed by TEM. Cross section samples were prepared using FIB milling (at 30 kV and 65–9.4 nA). In these cases, the cell surface was previously protected by a cord of platinum deposited using a gas injection system. Three different pieces of equipment were used to analyze the samples by TEM, a JEOL 3000F transmission electron microscope with a field-emission gun, a JEOL JEM ARM200cF microscope, and a FEI Tecnai F30 transmission electron microscope. The chemical composition of the semiconductor layers was also studied by means of EDX analysis and scanning TEM (STEM). High resolution TEM (HRTEM) and Fast Fourier Transform (FFT) tool were used to investigate the interface between the Ga(In)As cap layer and the metallization as well as the crystalline quality of the semiconductor layer beneath the metallization, prior and after the ALT.

The failure analysis is divided in three sections. Section 4.1 describes the condition of the solar cells before the ALT, Section 4.2 describes the effect of temperature on the degradation process by analysing the results obtained on the "reference" solar cells, which were only stressed with temperature in the climatic chambers, and finally, Section 4.3 describes the combined effect of temperature and current injection by analysing the solar cells with current injection in the climatic chambers.

## 4.1 Condition of the solar cells before the ALT

The characterization of solar cells before ALT showed that their efficiencies were of 35% around 500×. Figure 2 (top right) gathers the typical EQE curves for top, middle and bottom subcells. EL images for top and middle subcells can be seen in Figure 8. The bottom subcell is not shown because Ge has indirect gap and our EL setup is not able to detect its weak EL signal. Both the EQE (Figure 2 top right), and the EL (Figure 8) measurements show no anomalous behavior neither in the semiconductor structure nor in the metallization, implying that the three subcells and the two tunnel junctions are working properly and also that current is homogenously distributed.

A chemical and optical analysis of the solar cell surface, busbar and front metal fingers was performed. SEM has been used to explore the solar cell surface, including the front metallization grid. The EDX analysis in several points of the solar cell surface is shown in Figure 9. The compositional analysis from the spectra taken at several points on the cell with and without antireflective coating (ARC), are all gathered in Table 2. It is important to mention that EDX has a depth analysis of a few microns. No contaminants except for adventitious carbon are detected in this analysis, and the metallization grid and the solar cell active area show a smooth surface with no evidences of any mechanical defect.

The composition at the surface of the active area (point 4) is Al–Ga–In–P, suggesting an Al(Ga)InP window on the GaInP top subcell. The ARC, which covers the active area, the fingers (point 5) and the perimeter zone of the busbar surface (Point 6), is very probably made up of $TiO_x/AlO_x$. Points 1, 2 and 3 correspond to the metallic busbar not covered by the ARC, showing that the main constituents are Ag and Au. Analyses of metals covered by the ARC in points 5 (finger) and 6 (busbar) corroborate the composition of the metallization grid.

In order to achieve a deeper knowledge of the metallization grid state, cross-section pictures and chemical analysis have been performed after FIB milling. In spite of the uniform and compact metallization aspect along the whole thickness, some voids in busbar and fingers have been revealed in these cross-section pictures (Figure 10 top).

To investigate the homogeneity of the busbar composition, several line EDX analyses along the busbar thickness have been performed. The one shown in Figure 10 bottom was made at the left of the void seen in Figure 10 top where there is no ARC. It confirms Ag as the main metallization component which is sandwiched between two thin Au-based films, with a total Ag thickness a bit higher than 2 microns. Ga and In profiles are included in order to mark the Ga(In)As cap layer position.

Some other mapping analyses have been done along finger sections. Metallization follows the same scheme as in the busbars. In some analysis, some non-uniform thickness cross sections have been revealed after FIB milling, where a void in Ag bulk can be observed. The existence of these metal voids together with the surface roughness (see Figure 10 top) would be compatible with an electroplated silver layer.

STEM images on the metal finger cross-section shows clean and clear interfaces of metals and cap layer (Figure 11 top and bottom left). A more detailed EDX analysis made in the points marked in Figure 11 bottom left evidence the presence of Cl and S trapped as unexpected contaminants inside the silver metallization (see tables on Figure 11 bottom right). This fact can be explained since S and Cl are typical elements of most silver electroplating solutions. Regarding the gold contact on the Ga(In)As cap layer, it exhibits a very thin, clean and clear interface with a good crystallinity of the Ga(In)As cap layer (Figure 11, right).

Another zone of interest to be characterized on UMM solar cells is the buffer layer between the bottom cell and middle cell (see figure 1) where defects could create or propagate under given stress conditions. Accordingly, STEM images were carried out on the buffer layer cross section and no defects were detected (see Figure 12).

## 4.2 Influence of temperature in the degradation process

As explained in Section 2, each climatic chamber had two reference solar cells with no current injection for the duration of the test (see Table 1). One of the two cells per chamber was only stressed by the temperature of the chamber, while the other cell was also stressed by temperature but its dark I-V curve was measured periodically (same periodicity as for the solar cells under current bias). The purpose of these reference solar cells is to observe if any failure mechanism is activated by temperature soak alone, or by a combination of temperature soak and the bias applied during the measurement of the I-V curves. Figure 13 shows the dark I-V curves of the reference solar cells.

The left column of Figure 13 shows that the solar cells with temperature soak alone ("Ref_Low", "Ref_Mid" and "Ref_High") experience two effects: a) a decrease of the shunt resistance and b) a rise of series resistance which increases with temperature. This impact of the temperature was not observed in commercial lattice-matched GaInP/Ga(In)As/Ge triple junction solar cells [3]. An interesting feature is observed when comparing the left (temperature soak only) with the right column of Figure 13, where the time evolution of the dark I-V curve of solar cells stressed by temperature but with periodical measurement of their dark I-V curve is presented: the deleterious effect of series resistance caused by the temperature soak in "Ref_IV_Low" and "Ref_IV_Mid" is counterbalanced by the current injection when measuring the dark I-V curve. Just for the "Ref_IV_High" solar cell, current injection is not able to completely counteract its series resistance increase while the shunt resistance is even better at the end of the process (red dotted line) than that of the initial stage (black dotted line). Furthermore, an increase of recombination current with test time is found for the three cells "Ref_IV_Low", "Ref_IV_Mid" and "Ref_IV_High". The improvement of the shunt resistance could be related to the energy supplied by the current injection during the measurement process, that could reorder crystalline defects (such as in an annealing process), polarize vacancies, fill traps, etc. Therefore, the paths for current to leak substantially change due to current injection.

Figure 14 shows the illumination I-V curve of the same reference cells of Figure 13. In the left column the effects of temperature on both series and shunt resistance shown in Figure 13 are confirmed under illumination. An interesting aspect is that of the short circuit current variation. For "Ref_Low" and "Ref_Mid" solar cells, an $I_{sc}$ increase is detected which could be explained by the low shunt resistance observed in their dark I-V curves. This can be explained when the limiting subcell in the multijunction is shunted and then more current is allowed to flow through the solar cell as this subcell is under negative bias at $I_{sc}$. A more thorough explanation of this effect can be found elsewhere [20]. Since the "Ref_Low" cell exhibits a very low shunt resistance an artificial $I_{sc}$ increase comes up, while for the "Ref_Mid" cell the shunt resistance is higher, thus resulting in the typical tilt of the illumination I-V curve. Regarding "Ref_High" solar cell, which presents no shunt effect under illumination as expected from dark conditions, an $I_{sc}$ decrease is detected. This $I_{sc}$ reduction is confirmed in the right column of Figure 14 where no shunt current effect merges, as previously explained thanks to the beneficial effect of current injection. $I_{sc}$ decrease is more intense with the soak temperature.

In order to know the origin of the $I_{sc}$ reduction, we have measured the EQE of some cells. Figure 15 shows that "Ref_IV_Low" and "Ref_IV_High" cells exhibit degradation of their performance compared to the "Char_Ref" cell. For both cells, the top GaInP subcell exhibits the main degradation, followed by a very small degradation of middle Ga(In)As subcell. No significant degradation of the bottom subcell is detected.

In summary, temperature soak creates defects resulting in low shunt resistances, increase of series resistances, and increase of recombination current. The current injection during the dark I-V measurement has a beneficial effect by totally recovering the shunt resistance issues created by temperature although it is not able neither to completely recover the series resistance increase nor reduce the increase of recombination current which are activated by temperature alone. Furthermore, a decrease in $I_{sc}$ which increases with temperature is detected mainly because of the GaInP top subcell degradation. In the end, both reference cells at 165 °C, namely "Ref_High" and "Ref_IV_High" exhibit a power loss greater than 10% when characterized. Besides, a light decrease of $V_{oc}$ is observed on all solar cells as consequence of the recombination current increase. Therefore, the $FF$ decrease is the result of the variation of shunt and series resistances together with the $I_{sc}$ and $V_{oc}$ losses. We anticipate that temperature soak alone is enough to cause failures in the same time period as the solar cells with current injection, as will be shown in Section 4.3.

## 4.3 Combined influence of temperature and current injection in the degradation process

This section describes the evolution of the solar cells that were subjected to both temperature and forward bias, thus emulating nominal working conditions. Figure 16 shows the dark I-V curves and 500× illumination I-V curves of two failed cells at medium and high temperatures, namely, "Fail_Mid" and "Fail_High" cells. No figure with a cell failed at low temperature is shown, since as we have stated in Section 2, the duration of the degradation test was not long enough to achieve failures at low temperature. Dark I-V curves clearly show: a) no effect of shunt resistance, b) increase of recombination current and c) increase of series resistance, thus confirming what described in Section 4.2 for the solar cells without current injection. Illumination I-V curves show: a) increase of series resistance affecting $FF$, b) decrease of $I_{sc}$ and c) small drop of $V_{oc}$ Therefore, the increase in series resistance together with $I_{sc}$ loss are the failure modes that dominate the aging process leading to the reliability values of Section 3 and Figure 7. Accordingly, we evaluate below the origin of these failure modes.

In order to find out the origin of the series resistance increase four wire measurements were carried out between two points of the busbar to obtain the front contact resistance of several failed solar cells and also of not tested ("Char_Ref", see Table 1) solar cell. Figure 17 shows the illumination power loss (at 500×) vs. front contact resistance in which the relationship between the power loss and the increase in grid resistance is linear with a coefficient of determination higher than 0.99 for all degraded solar cells in zone 3 (see Figure 4 for the meaning of the "zones"). All measurements produced similar results (i.e. linear relation between measured resistance and power loss) regardless of the points of the busbar chosen to measure the resistance in between. The linear relation between illumination power loss and front contact resistance suggests that failures are mainly caused by the degradation of the front contact. Figure 17 shows the interesting fact that not only the failed cells (power loss higher than 10%) follow this relationship but also the cell degraded but not still failed, namely the "Deg_High_Rs" cell. In these cases, the solar cells were at zone 3 of Figure 4 on which series resistance dominates. However, the "Deg_Low_Rec" cells fall out of the red line of Figure 17 since those cells were at the zone 1 of Figure 4 where recombination current increase rules degradation while series resistance has not still appeared.

Solar cell degradation has been simulated using the same three dimensional (3D) distributed model applied in Section 2 to determine the current to inject in the cells in order to emulate the nominal photo-current conditions at 500×, while avoiding electrical overstress that would produce artificial failures [21]. Accordingly, the experimental dark I-V curve of the "Char Ref" which is a reference solar cell that has not been stressed at all (see Table 1), together with that of the "Fail High" solar cell have been fitted in Figure 18, while the corresponding values of $J_{01}$, $J_{02}$, shunt and series resistances for each subcell are presented in Table 3. The initial $J_{01}$ and $J_{02}$ values chosen were the commonly reported ones [22] for the top (GaInP) and bottom (Ge) subcells. For the middle subcell (GaInAs) the initial values were estimated from its bandgap. Afterwards, more accurate values were obtained by fitting the dark I-V curve. The $J_{01}$, $J_{02}$ parameters of the "Fail High" cell after degradation are almost impossible to pin down for each subcell from the whole multijunction dark I-V curve. In fact, a good fit of the I-V curve can be achieved using different $J_{01}$, $J_{02}$ combinations. In order to have a first insight about the degradation of the subcells, the same $J_{01}$ and $J_{02}$ increment factor was applied to each subcell except for $J_{02B}$ since it is well known that there is no contribution of $J_{02}$ for Ge subcells [23]. A good fit was achieved by using an increment factor of 1.5. Therefore, we are assuming that the three subcells have been degraded although their extent will be determined by EQE. Metal sheet resistances were calculated from the front contact resistance values obtained from the previous four wire I-V measurements (see Figure 17), resulting in values for the metal sheet resistance of 2.65·10$^{-2}$ and 5.3 10$^{-2}$ Ω/□ for the "Char Ref" and "Fail High" solar cells, respectively.

Figure 18 shows that the "Char Ref" simulation fits very well the experimental curve. Regarding the "Fail High" solar cell simulation, a very good fit was achieved by both reducing the shunt resistance and increasing 1.5 times the $J_{01T}$, $J_{01M}$, $J_{01B}$, $J_{02T}$, $J_{02M}$ values (see Table 3). But, when trying to simulate the series resistance effect of the "Fail High" cell by only increasing the value of the metal sheet resistance (2.65·10$^{-2}$ → 5.3 10$^{-2}$ Ω/□) resulted in an almost negligible impact on the dark I-V curve (not shown in Figure 18 for the sake of simplicity). Therefore, the specific front contact resistance was increased until the "Fail High" simulated curve fits the experimental one, which happened for a specific front contact resistance around 3.5·10$^{-3}$ Ω·cm$^2$ (i.e. much higher than the typical values around 5·10$^{-4}$-10$^{-5}$ Ω·cm$^2$ of the non-degraded one).

In order to give physical support to these simulations, EL analysis of the same two cells was carried out. Firstly, we measured the GaInP top subcell EL of both cells (Figure 19 top) and after that, we simulated the EL patterns of both cells with our 3D model (Figure 19 middle). As it can be seen, there is a very good fit between the experimental and simulated EL patterns when the parameters of Table 3 are used (including the estimated value of the specific contact resistance of 3.5·10$^{-3}$ Ω·cm$^2$). It can be argued that a very good fit of the "Fail High" solar cell (red curve of Figure 18) could also be achieved by using different values of some components of the front series resistance. For example, we have found that by increasing only the metal sheet resistance around 40 times ($2.65 \cdot 10^{-2}$ Ω/□ → 1 Ω/□), a good fit of the red curve is also achieved. However, if a value of 1 Ω/□ is used for the metal sheet resistance, the simulated EL pattern of Figure 19 bottom is achieved. This pattern is significantly different from the experimental one (Figure 19 top right) discarding a failure dominated by a degraded metal sheet resistance. This last simulated pattern results from the fact that the solar cell only has bond wires at the right and left sides of the busbar, producing further areas darker as increasing the metal sheet resistance in contrast with the "symmetrical rounded" experimental pattern. Consequently, we can conclude that a degradation of both the specific contact resistance and the metal sheet resistance is taking place, so we want to know the origin of both degradations (although the metal sheet resistance has an almost negligible influence on the solar cell degradation).

In order to know the origin of the front metal grid degradation, solar cells have been examined in the same way as described in section 4.1. In this case, metallization grid of the "Fail High" solar cell, both in busbar and fingers, shows a blistered surface (Figure 20), in contrast with the smooth surface seen in the "Char_Ref" solar cell (Figure 9). EDX analysis was carried out in several points and the compositional analysis from the spectra can be seen in Table 4.

Some remarkable differences should be pointed out when comparing Table 4 with Table 2 of the "Char_Ref" solar cell, which accounts for the situation before degradation (section 4.1). Again, points 1, 2, 3 in Table 4 were taken in the metallic area not covered by the ARC, while 4, 5, 6 correspond to the active area, a finger and a part of the busbar covered by the ARC, respectively. The amount of C on the surface has increased a lot in this failed cell. The presence of a new element, namely, Si, is now evident in all the analysed points. Its origin is probably related to its evaporation from the thermal paste used to join the Solar cell to the aluminium base plate (see Figure 1). Chlorine, which was also detected in the "Char Ref" cells, is only absent in the solar cell active surface (point 4) while S is only present in the busbar surface not covered by ARC. In summary, silicon is almost everywhere while both chlorine and sulphur are in the metal grid. Chlorine is at any grid point of the metal grid while sulphur is only in the uncovered grid (exposed to ambient conditions). The roughness increase of the metallization surface is more evident in the cross-section picture (Figure 21 top). A remarkable fact is that in the busbar area uncovered with ARC, the roughness is higher than at the covered parts including fingers. In Figure 21 bottom left, the roughness increase is visible in the form of small granules, whose composition seems to be an Ag-Cl and Ag-S mixture. Their sizes are in the order of microns. ARC seems to slow down the formation of these granules. This is probably due to the barrier effect that the ARC plays on top of the metallization, preventing the metallization constituents moving towards the top surface, where they can interact with air during the ALT. As a consequence of the movement of the metal grid constituents towards the surface, an increase in the void density of the front metal grid has been detected in several cross-section images taken at different busbar and finger zones. Some of these voids appear now partially filled (Figure 21 bottom right). The EDX analysis of these stuffing voids shows a huge contamination of chlorine together with a silver depletion around the void. It should be reminded that the presence of Cl and S was already detected in the "Char_Ref" cells of Section 4.1 but the difference is that now Cl and S have moved and concentrated in several regions where they form compounds of silver that poison the electric performance of the front metal grid. All these effects can explain the two times reduction of the front contact resistance of the cells subjected to temperature and current injection degradation.

Now we are going to analyse the origin of the specific contact resistance increase (see Table 3). Figure 22 (top left) shows a clear evolution of the metal finger cross section of a "Fail_High" solar cell when compared with that of "Char-Ref" cell (see Figure 11). Now the interfaces between metals and also between metals and cap layer are not abrupt but very diffused (Figure 22 top left). Besides, the EDX analysis of gold layer shows sulphur and silver contamination (Figure 22 bottom left). Additionally, there are intrusions of gold, sulphur and silver in the Ga(In)As cap layer which modify the properties of the specific front contact resistance. A deeper insight by HRTEM image of the Ga(In)As cap layer and the metallization interface (Figure 22 top right), shows a deterioration of the semiconductor layer crystalline quality, which gives rise to the different contrast fringes observed in the HRTEM image. The FFT image (Figure 22 bottom right) confirms this point, showing extra spots disturbing the cubic symmetry of the Ga(In)As layer. Therefore, we can conclude that the Au/Ga(In)As cap layer interface is much degraded in the "Fail_High" cell and consequently, the specific front contact resistance exhibits a high increase.

Finally, the other origin for the solar cell failure, namely, the decrease of the short circuit current was investigated. When measuring the EQE of "Fail_High" solar cells, we were unable to measure the top and middle subcells because of shunt problems. The EQE of the Ge bottom subcell was properly measured showing a mid-degradation at a lower extent than that experienced by the top and middle subcells shown in Figure 15. Since the short-circuit current decrease here (Figure 16) is similar to that experienced by the solar cells with temperature soak alone (Figure 14) which was due to the top and in a lesser extent to the middle subcell degradation (Figure 15), we can assume that the degradation now is also mainly due to the top and in a lesser extent to the middle subcell. The degradation of the EQE can be associated with the increase of the recombination current shown in Figure 4.

In summary, the time evolution of the degradation process would be: 1) increase of the recombination currents caused by temperature stress that reduces the EQE mainly of top and middle subcells with the subsequent short circuit current decrease, 2) as time goes by, an increase of series resistance caused by both temperature and current injection is experienced by the cells as a result of the specific front contact resistance degradation. An important remark is that since in this work we have stated a failure criterion of 10%, the origin of failure is mainly the series resistance increase together with a short circuit current decrease. However, with a stricter failure criterion, namely 5% or below, the main cause of failure would be the increase of recombination current (Zone 1 of Figure 4) that impacts on the decrease of the short circuit current, open circuit voltage and fill factor.

## 5. CONCLUSIONS

In this paper, the statistical reliability analysis and the failure analysis after a temperature ALT on upright metamorphic $Ga_{0.37}In_{0.63}P/Ga_{0.83}In_{0.17}As/Ge$ triple-junction solar cells were carried out. By assuming a failure criterion of 10% power loss at 500×, an Arrhenius activation energy of 1.39 eV was determined. Accordingly, a warranty time of 72 years for a place like Tucson (AZ, USA), was determined which is an upper time limit that will be reduced when considering additional stress factors such as humidity, thermal cycles, etc.

Regarding the failure analysis, we have detected that temperature soak alone (without current injection) is enough to degrade the cell performance by decreasing the shunt resistance and increasing both series resistance and recombination currents. This results in: a) decrease of $I_{sc}$ (which increases with temperature) mainly because of GaInP top subcell degradation b) a light reduction of $V_{oc}$ and c) a $FF$ fall mainly due to a resistive loss caused by a severely degraded contact layer. This negative impact of temperature soak was not detected in lattice matched GaInP/Ga(In)As/Ge triple junction solar cells.

When solar cells were forward biased with the current injection equivalent to the nominal working conditions at 500× by keeping the temperature soak, an increase of series resistance together with a reduction of short circuit current is detected. This forward biasing does not produce an additional reduction of the short circuit current. However, the series resistance is increased by the forward bias. The failure analysis of the series resistance shows that: a) a lot of voids in the silver layer of the front metal grid appear which in some cases are stuffed with chlorine compounds together with a silver depletion around the void. Also, Cl and S concentrate in several regions where they form compounds of silver that poison the electric performance of the front metal grid. All these effects explain the two times reduction of the front contact resistance. b) The Au/Ga(In)As cap layer interface is greatly degraded because of the contamination by sulphur and silver of the gold layer. Additionally, there are intrusions of gold, sulphur and silver in the Ga(In)As cap layer which modify the properties of the specific front contact resistance. There is also a deterioration of the Ga(In)As cap layer crystalline quality evidenced by the disturbance of its cubic symmetry. All these effects result in a huge increase of the specific front contact resistance. The problems associated with the current front contact could be circumvented by using new generation of metallic contacts. However, the increase of recombination currents (and their impact on the cell performance) should be solved. Finally, it must be stated that the entirety of results of this paper are only applicable to the particular solar cells with their specific packaging analysed in this work although several tendencies could be extrapolated to similar devices.

## 6. ACKNOWLEDGEMENTS

The authors thank to Dr. Andreas Bett from FhG-ISE for supplying the UMM solar cells within the framework of the NGCPV project (European Commission funded with Grant agreement ID: 283798) and for his help and interesting comments during the NGCPV project. This work has been supported by the Spanish MINECO through the project TEC2014-54260-C3-1-P by the Comunidad de Madrid through the project MADRID-PV2 (S2018/EMT-4308) and by the Universidad Politécnica de Madrid with the project RP150910B12. Technical and scientific support for the TEM measurements was given by Dr. Javier García of the Centro Nacional de Microscopía Electrónica. Part of the microscopy works have been conducted in the "Laboratorio de Microscopias Avanzadas" at "Instituto de Nanociencia de Aragon - Universidad de Zaragoza". Authors acknowledge the LMA-INA for offering access to their instruments and expertise.

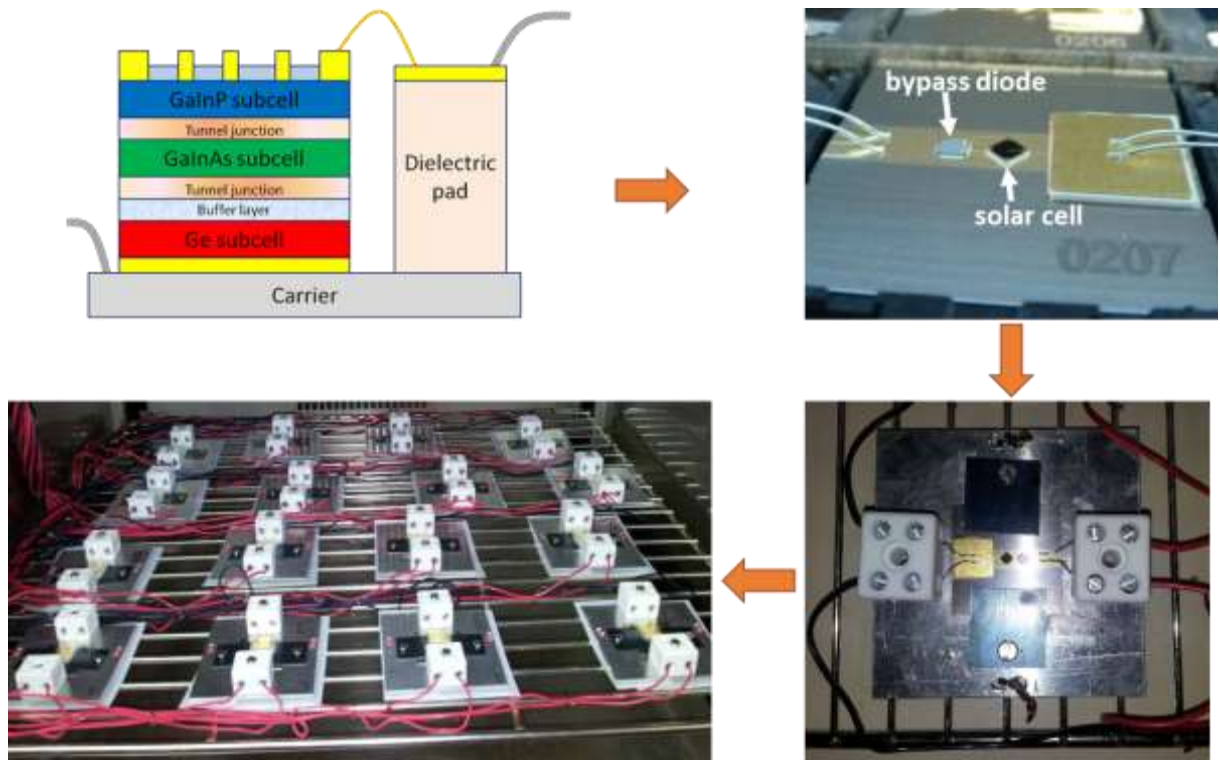

Figure 1. Schematics of the tested UMM CoC (top left), a picture of a CoC (top right), picture of the CoCs onto aluminum heatsinks with connectors (bottom right) and experimental setup of several CoCs inside one of the climatic chambers (bottom left). The bypass diodes such as those shown on the top right picture were disconnected from all the solar cells before the ALT

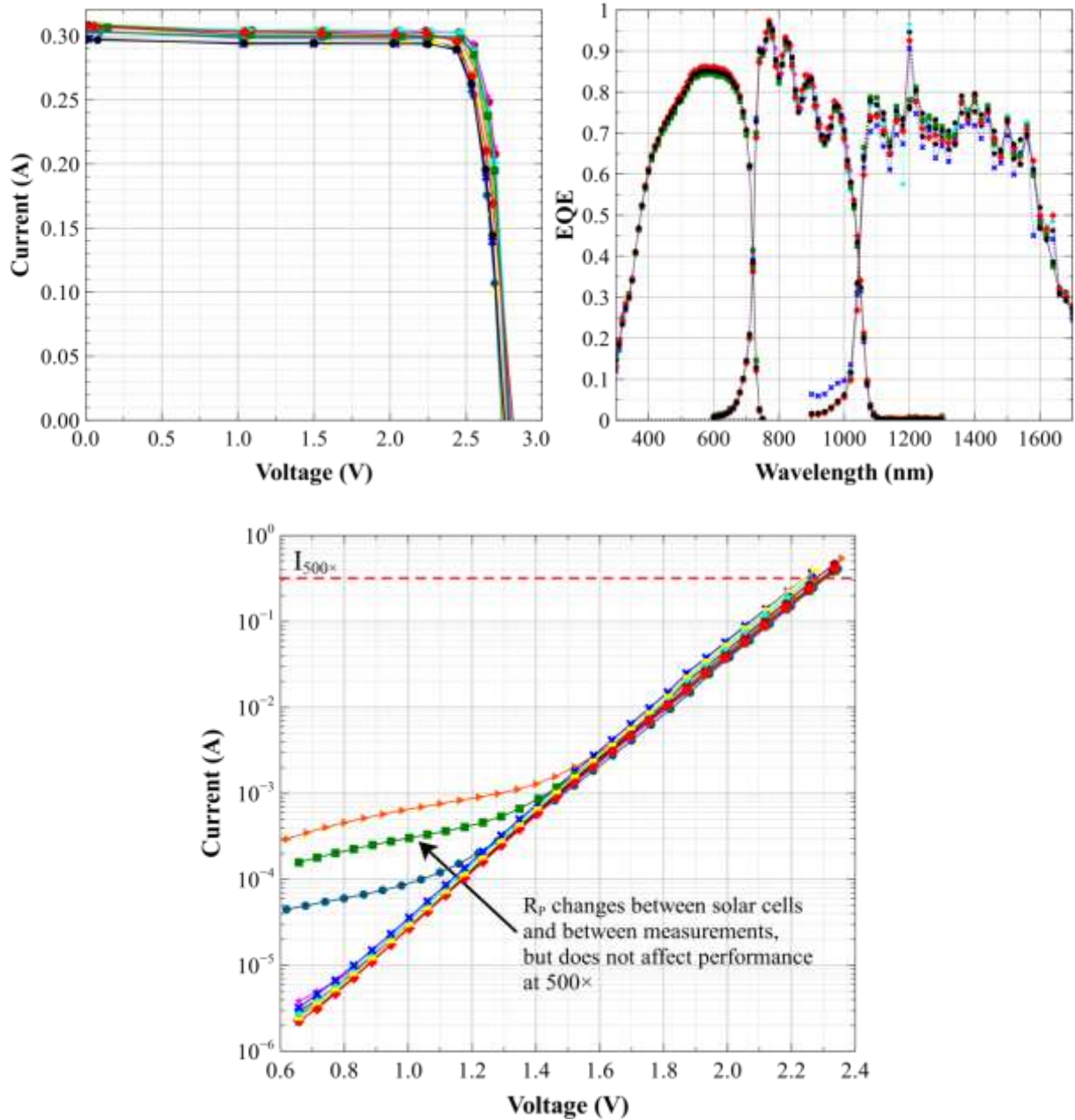

Figure 2. Illumination I-V curves at 500× (top left), EQE (top right), and dark I-V curves (bottom) of several solar cells before the ALT. A feature of these solar cells is that the shunt resistance changes with time (goes higher and lower), but it never gets low enough to affect performance at nominal working conditions.

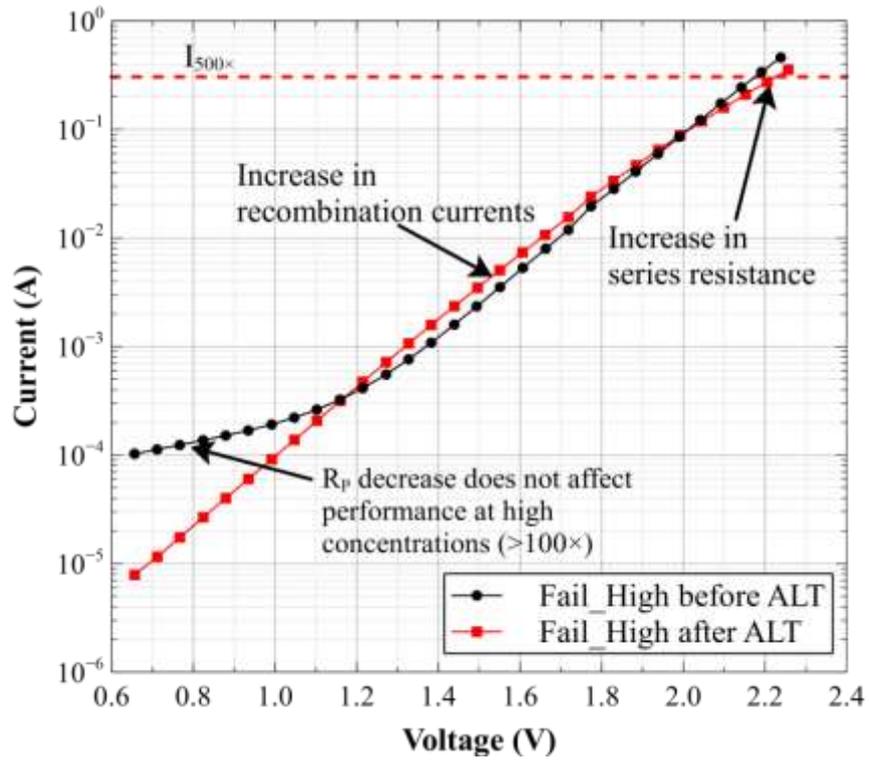

Figure 3. Typical dark I-V curve evolution of a degraded solar cell ("Fail_High" of Table 1). Two main degradation mechanisms are observed: (1) an increase in recombination current and (2) an increase in series resistance. The shunt resistance improves/worsens between cycles without affecting the performance of the solar cell at 500×

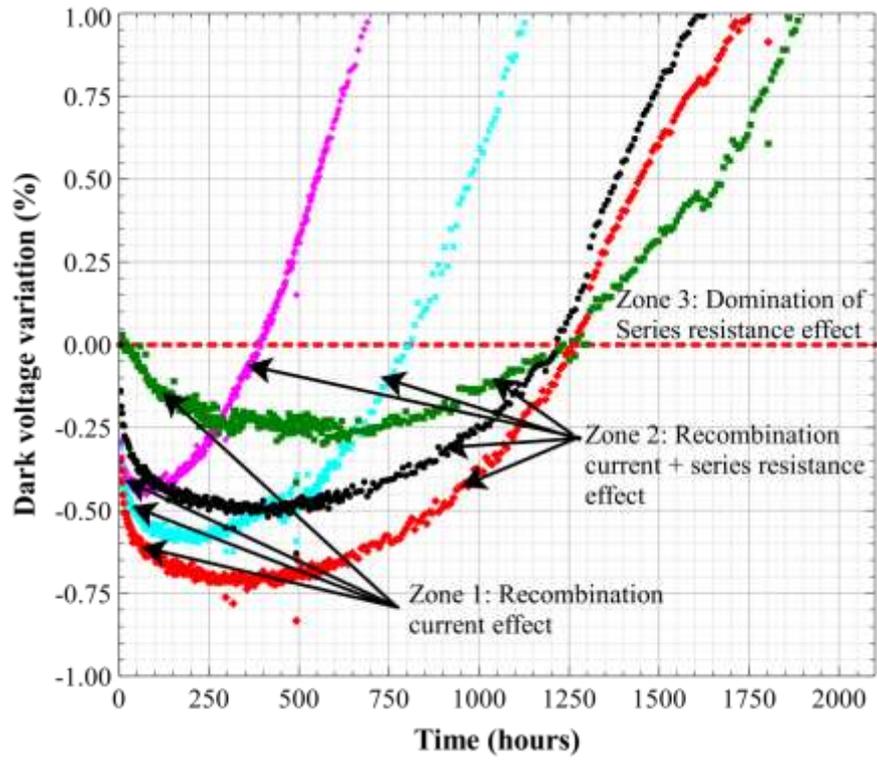

Figure 4. Evolution of the dark voltage at equivalent illumination operation conditions ($I_{500\times}$ of Figure 4) as a function of test time for several "Fail_High" solar cells. The time evolution shows that initially the solar cells exhibit a slight increase in recombination current (dark voltage decreases) while after some hundreds of hours the series resistance increase (dark voltage increases) dominates the degradation process. Different colors mean different devices

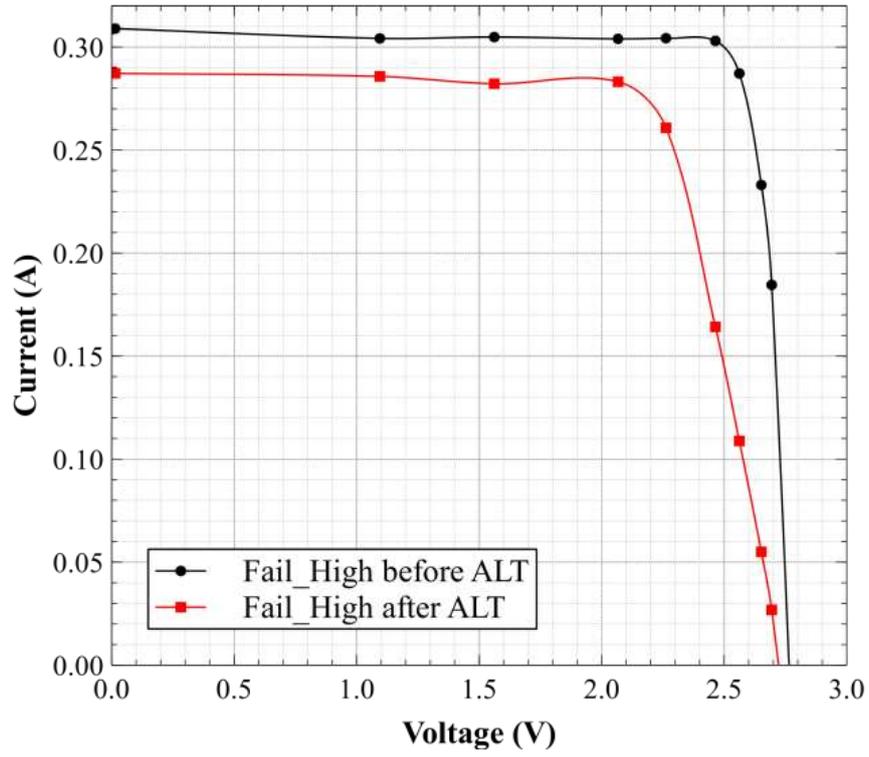

Figure 5. Comparison of illumination I-V curves at 500× before and after ALT for a solar cell degraded at 165ºC.

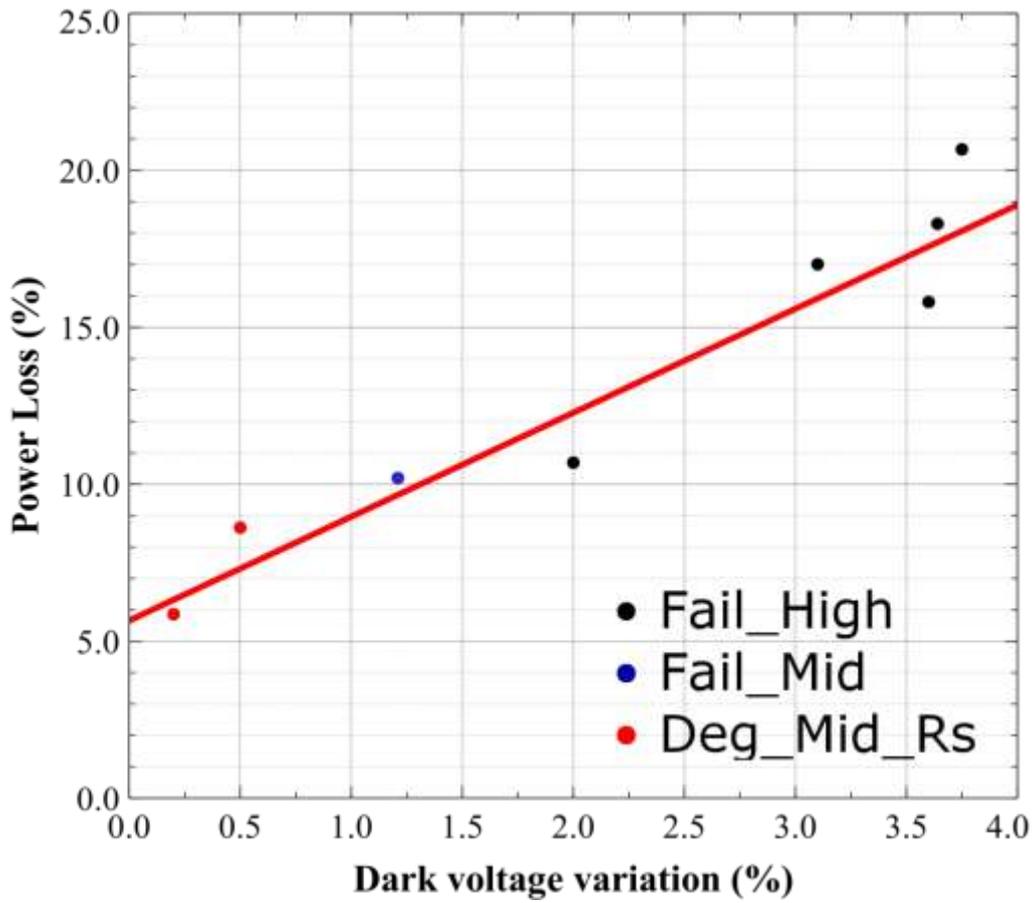

Figure 6. Experimental (black points) and average tendency (red line) illumination power loss at 500× vs. voltage variation in darkness (at the current equivalent to 500× illumination). Notice that at zero volts variation in darkness there is an illumination power loss of around 5% because of the increase in recombination current (decrease of dark voltage) although in terms of voltage variation is compensated by series resistance (increase of dark voltage) as can be seen in Fig. 4.

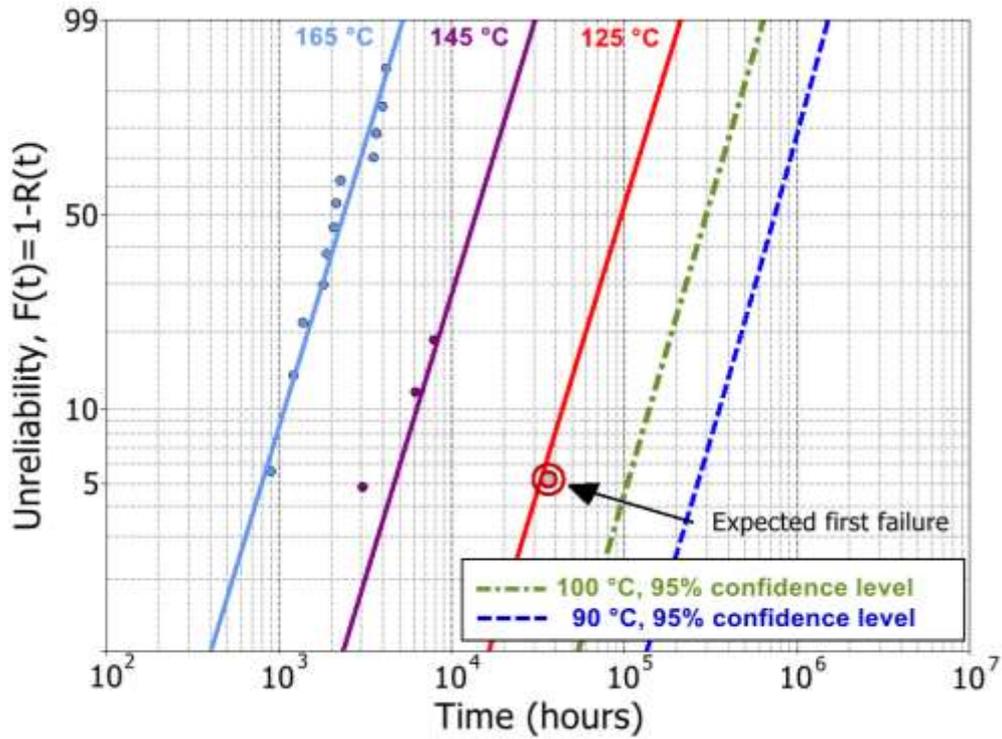

Figure 7. Unreliability vs. time plot for every stress temperature used in the ALT and extrapolated results at 90ºC and 100°C with a 95% confidence level. The dark red target represents the time for the expected first failure in the 125 °C test (around 40,000 hours)

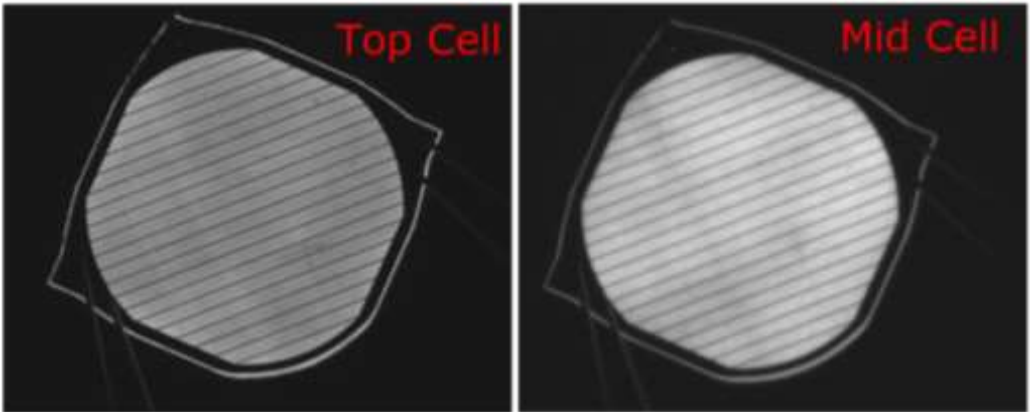
Figure 8. Typical EL maps of top and middle subcells before the ALT

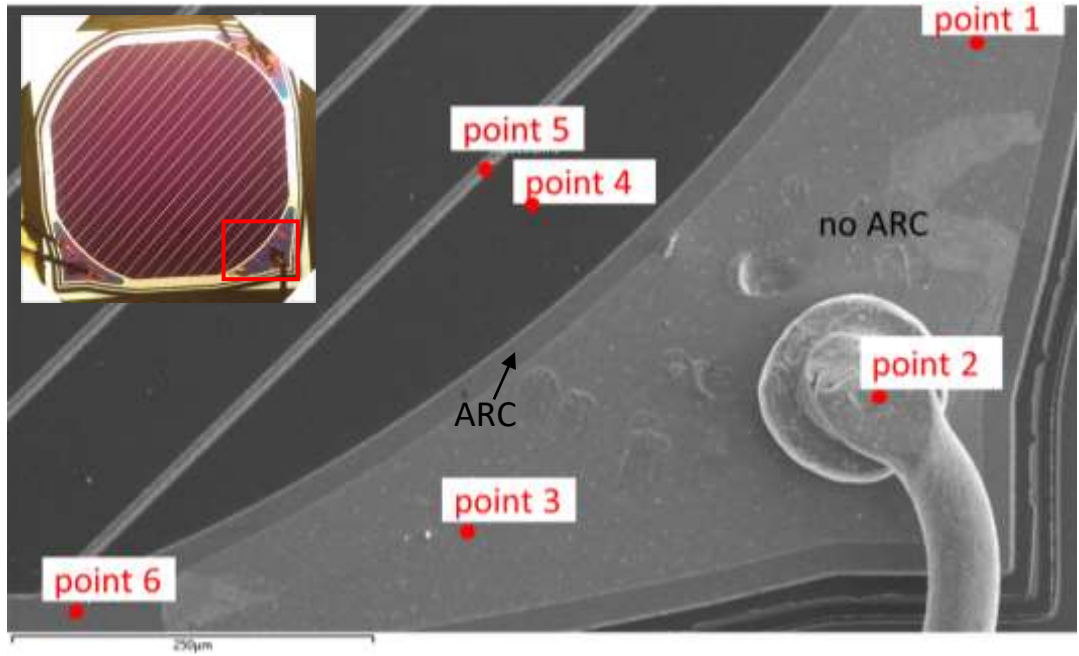

Figure 9. Planar view SEM image of the "Char_Ref" solar cell surface, showing the points where EDX analysis has been performed. The clearest area of the busbar has no ARC while out of this region, ARC is deposited. The inset on the top left shows the region of the solar cell being analyzed.

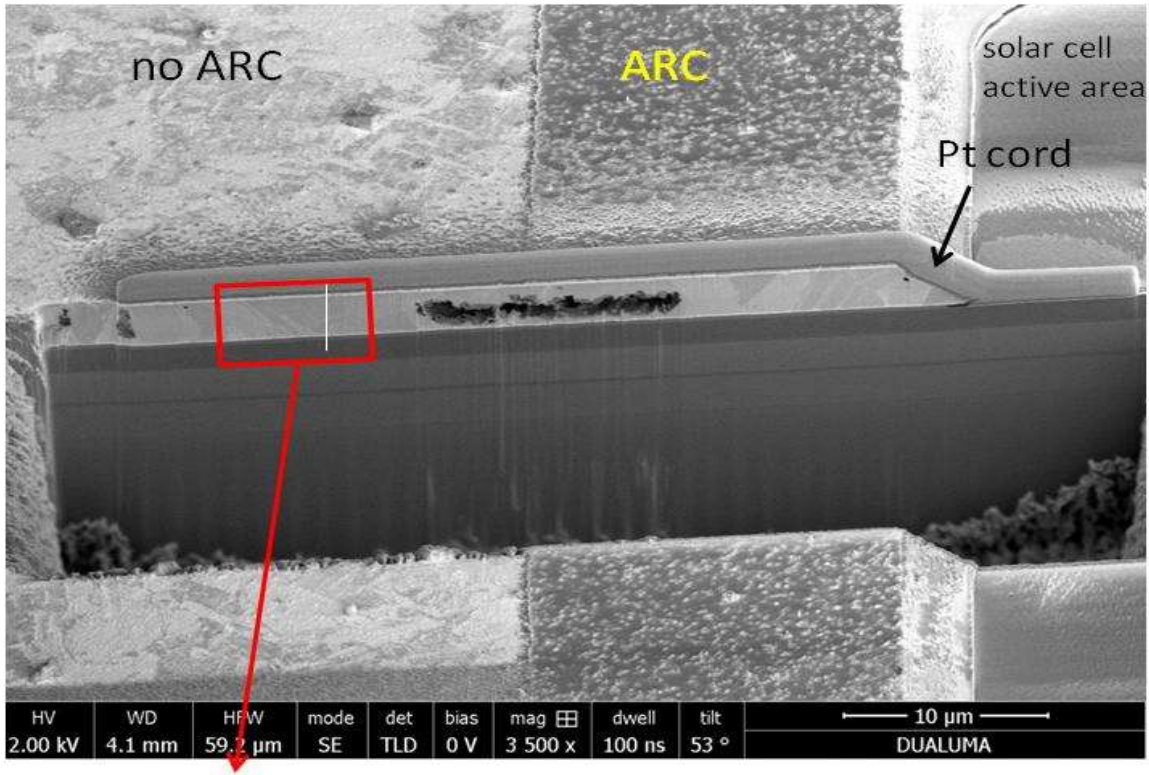
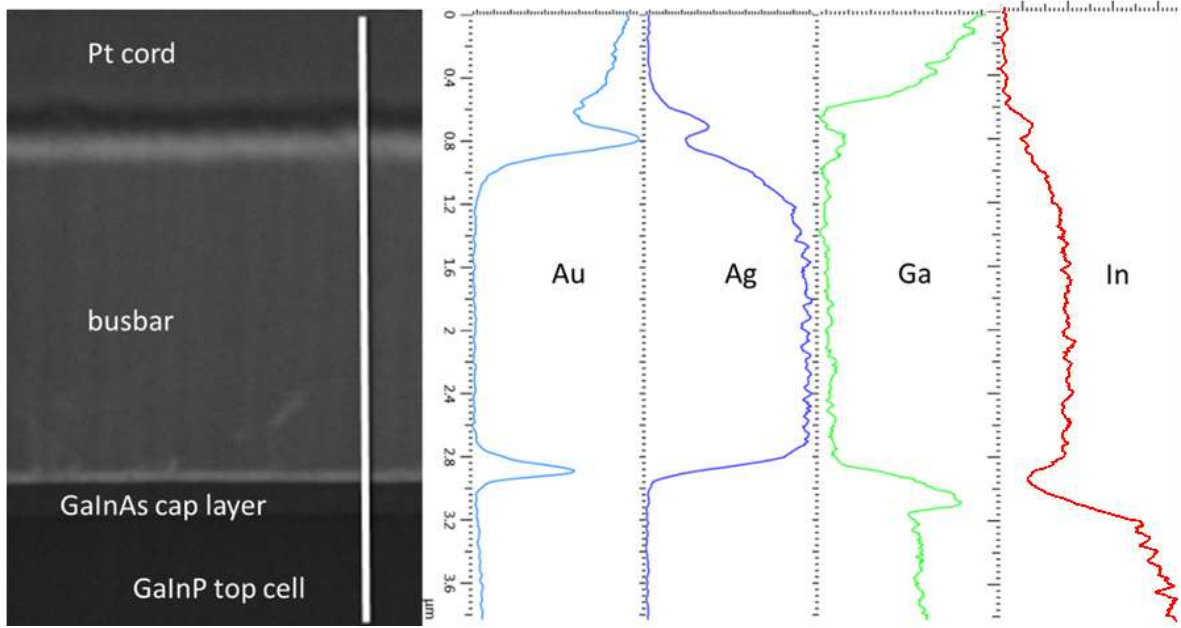

Figure 10. Top: SEM micrograph after FIB milling of the bus metallization cross-section of the "Char_Ref" solar cell. Pt cord is deposited in order to protect sample surface during FIB. Bottom: (left) SEM micrograph of the busbar between the Pt cord and the Ga(In)As cap layer. White vertical line corresponds to the length scale on the right. (Right) Main element profiles along the analyzed length. Au, Ag, Ga and In signals have fake artifacts around 0.6-0.7 microns depth due to Pt signal. Ga implantation in the Pt cord due to FIB procedure is the cause of the Ga increase in the first half micron.

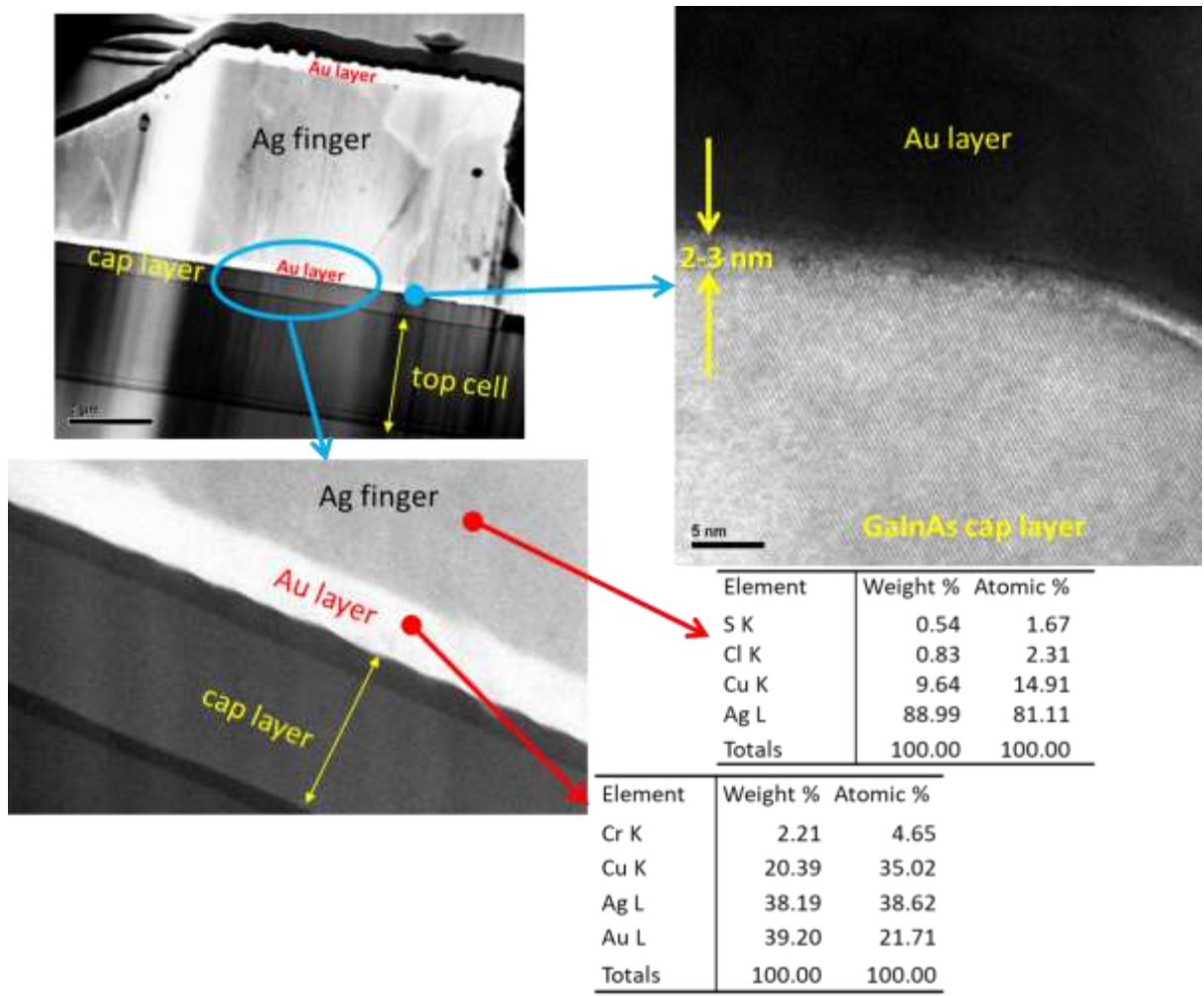

Figure 11. Cross-section view STEM image of a finger of the "Char_Ref" solar cell (top left), showing the area where EDX analysis has been performed (bottom left). Copper presence in the EDX analysis derives from the sample holder. HRTEM of the Au/Ga(In)As interface (right)

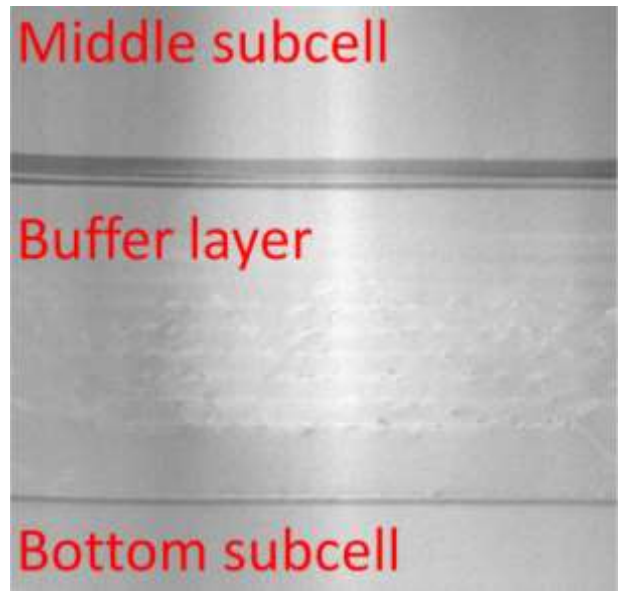

Figure 12. Cross-section view STEM image of the lower part of the semiconductor structure of a "Char_Ref" solar cell, highlighting the buffer layer.

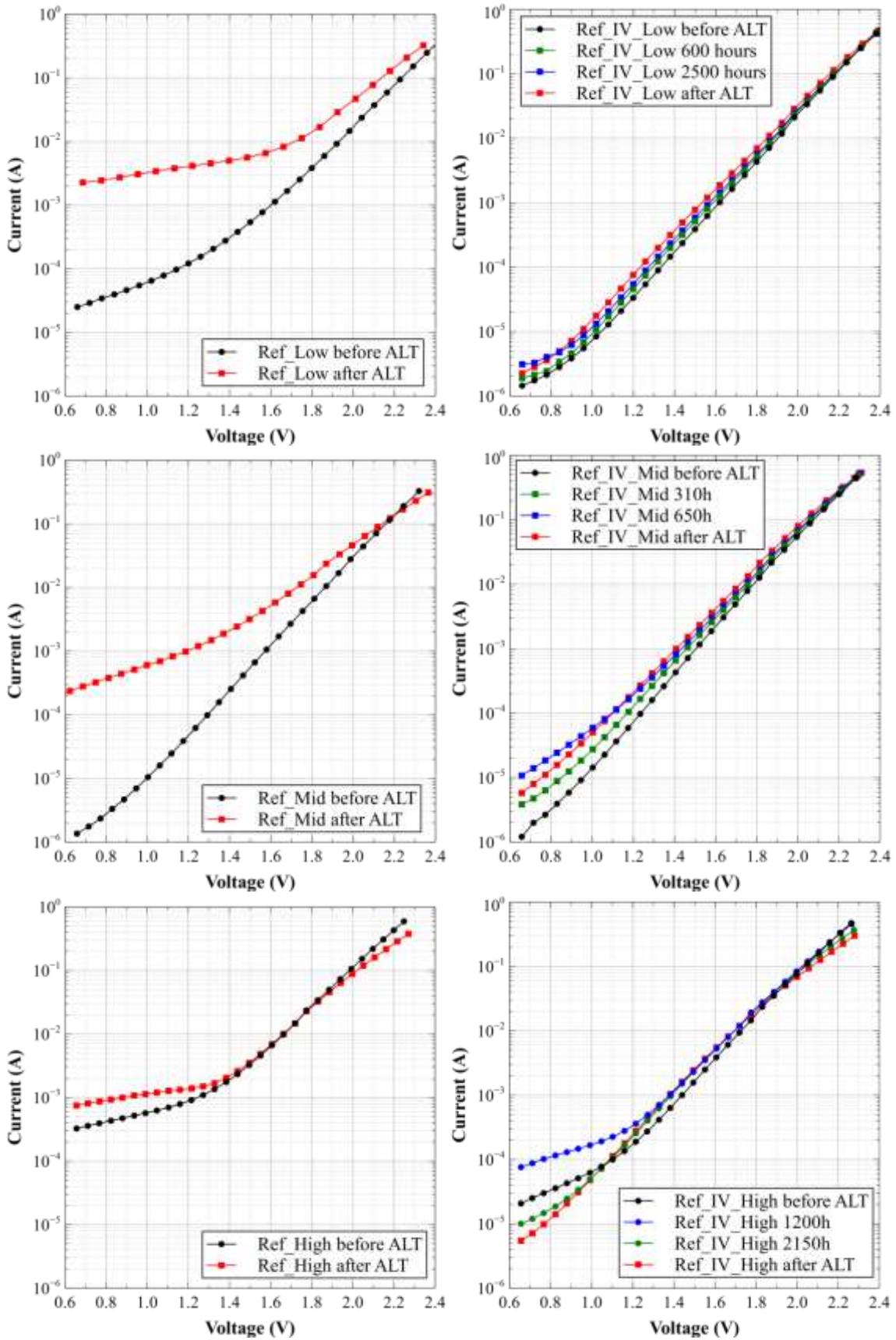

Figure 13. Comparison of the dark I-V curves of the reference solar cells (those without I-V measurements) before and after ALT (left). Evolution of the dark I-V curves for different times of the reference solar cells (those with I-V measurements) during ALT (right).

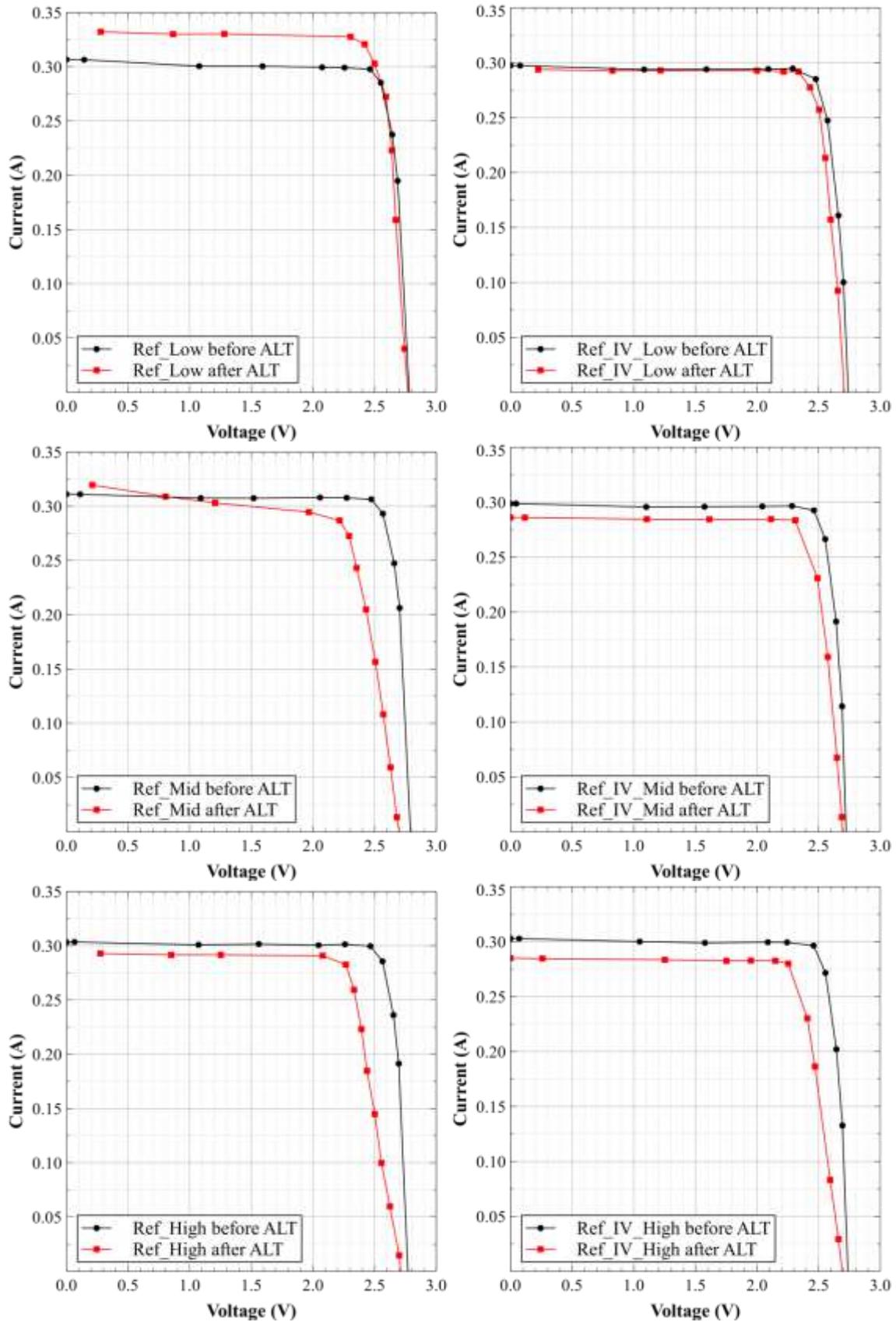

Figure 14. Comparison of the Illumination I-V curves at 500× of the reference solar cells of Figure 13 before and after ALT.

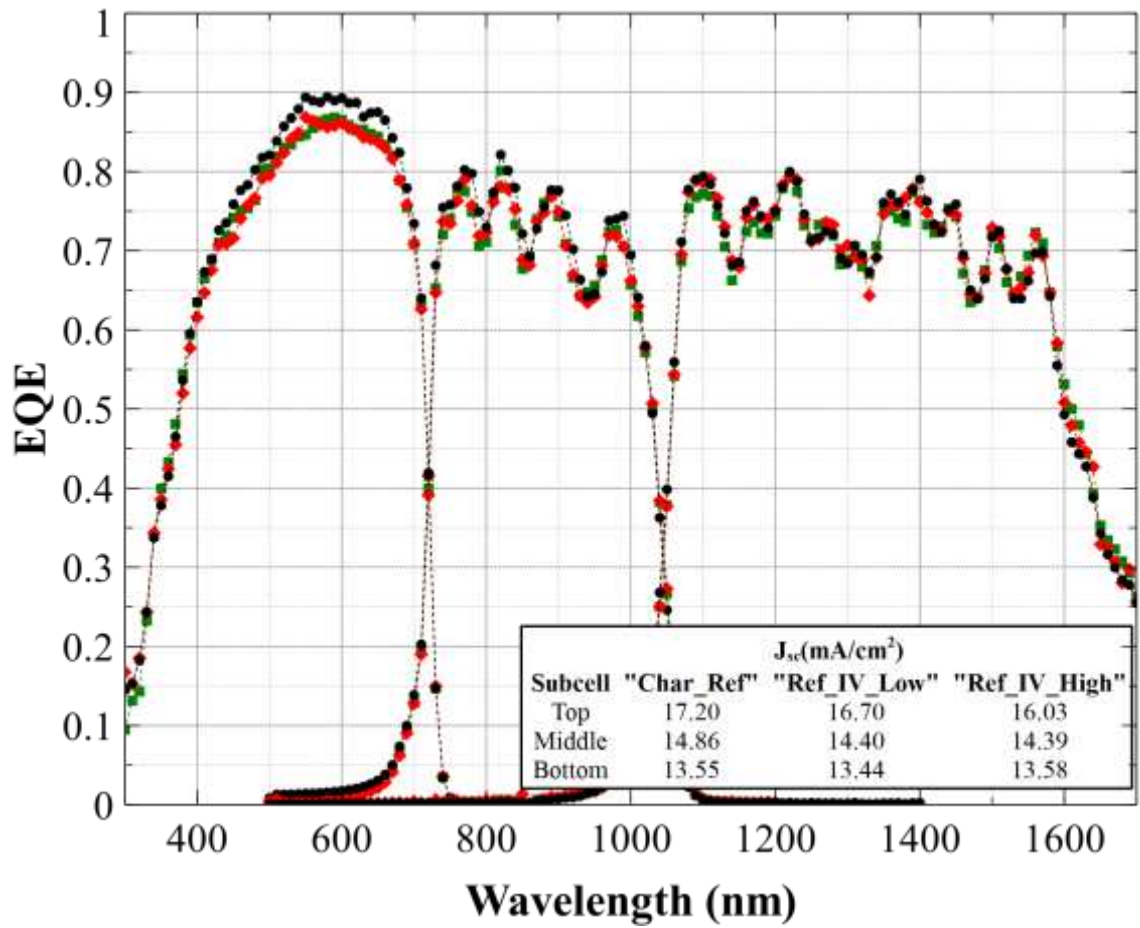

Figure 15. Comparison of the EQEs of "Char_Ref" (black circles), "Ref_IV_Low" (red circles) and "Ref_IV_High" (green circles) solar cells. The bottom inset shows the calculated $J_{sc}$ of each subcell under spectrum AM1.5d ASTM G173–03.

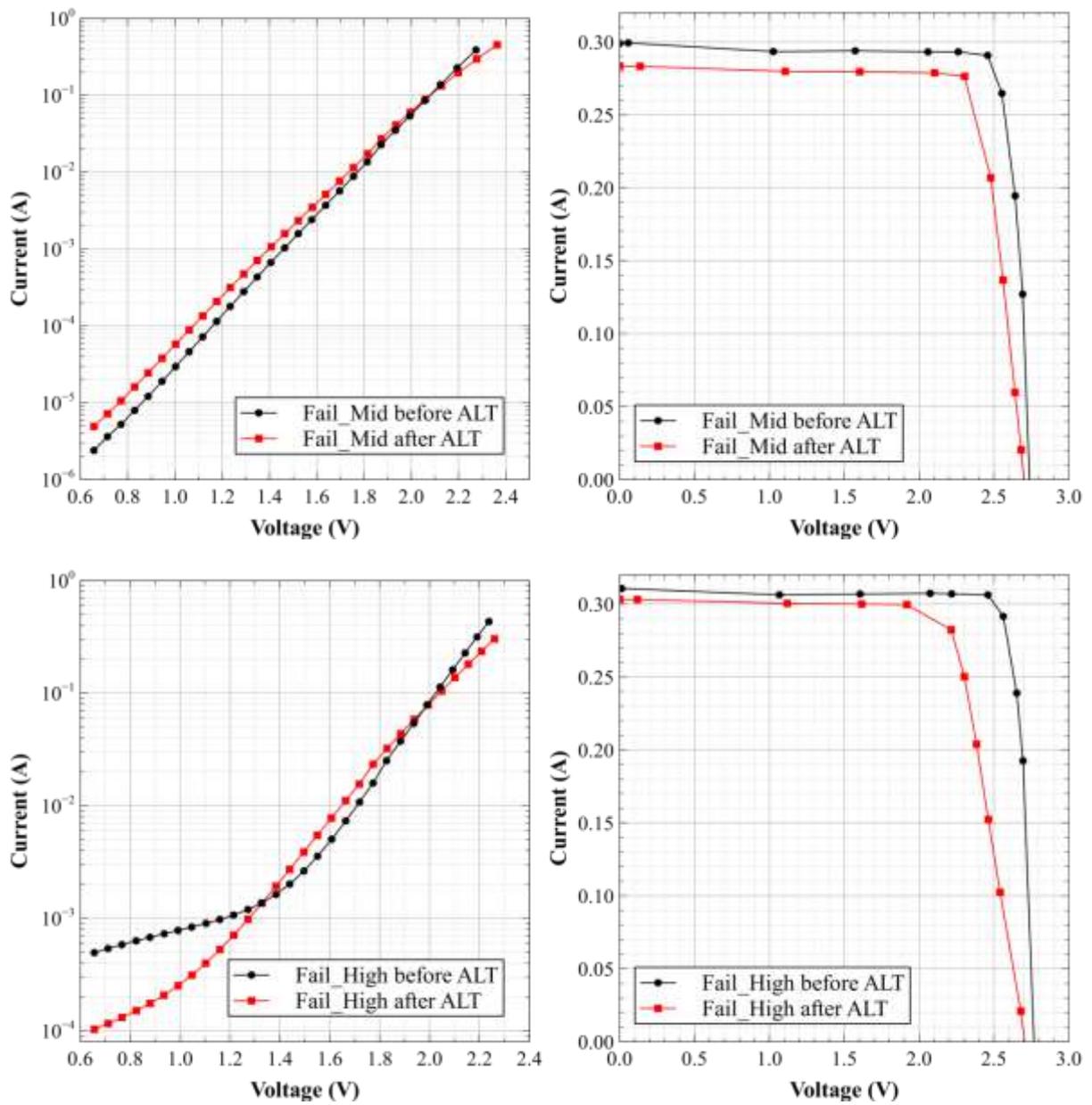

Figure 16. Dark I-V curves (left) and illumination I-V curves at 500× of solar cells with current injection before and after ALT. Both solar cells exhibit a power loss greater than 10%.

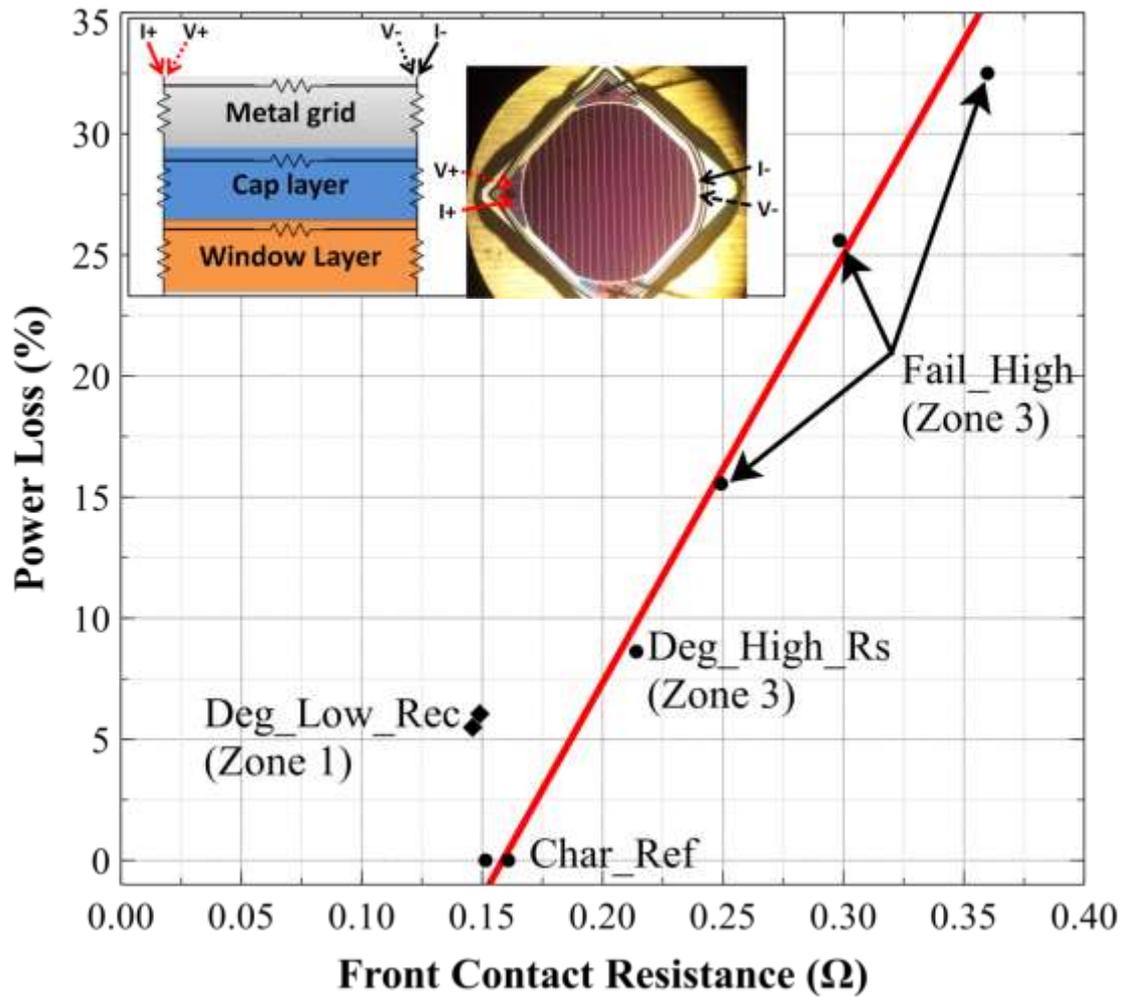

Figure 17. Illumination power loss vs. front contact resistance plot. Notice that front contact resistance measured includes the specific front contact resistance, grid resistance and lateral Ga(In)As cap layer resistance. Each dot represents a measured solar cell. The power loss was obtained comparing the light I-V curves at 500× before and after ALT. See Figure 4 for the meaning of Zone 1 and Zone 3.

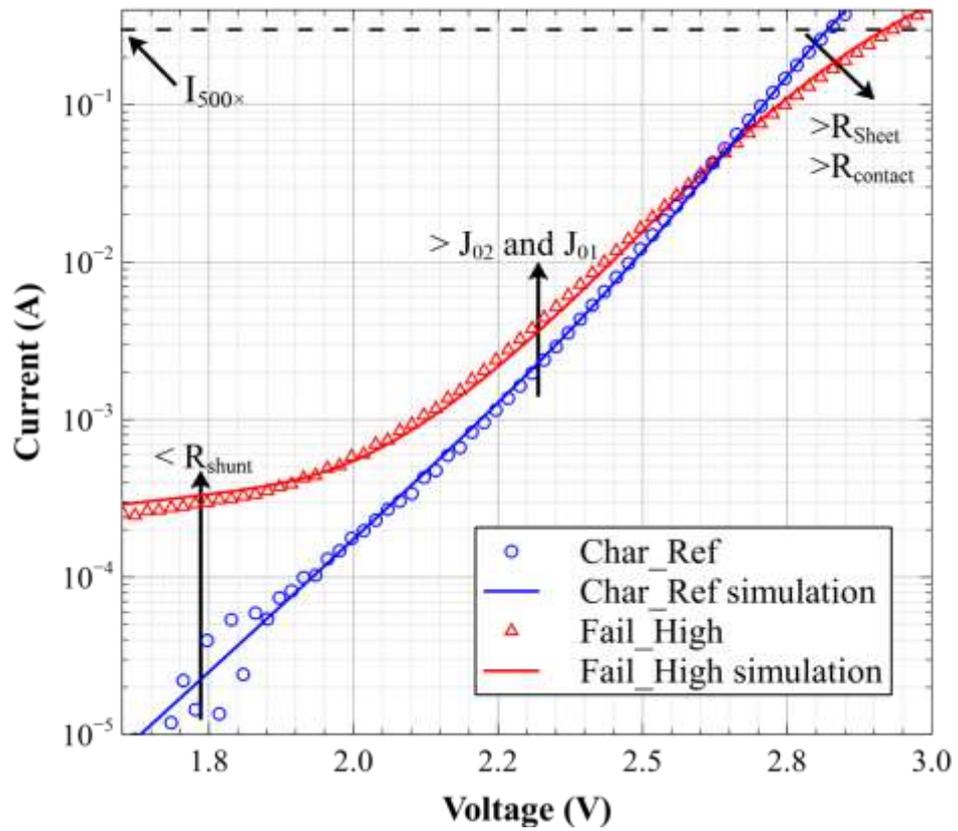

Figure 18. Experimental and fitted dark I-V curves of a non-degraded at all "Char Ref" and a failed "Fail High" solar cell. The fitting parameters used are presented in Table 3.

| GaInP Top Cell | "Char_Ref" solar cell | "Fail High" solar cell |
|---|---|---|
| Experimental | 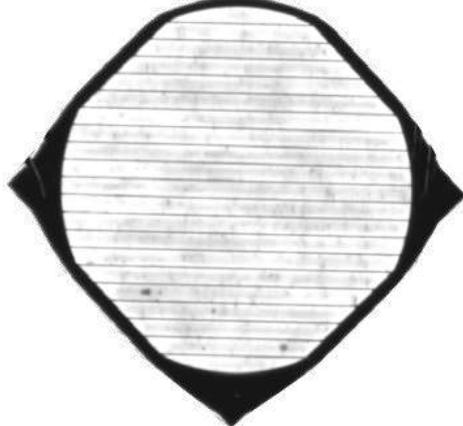 | 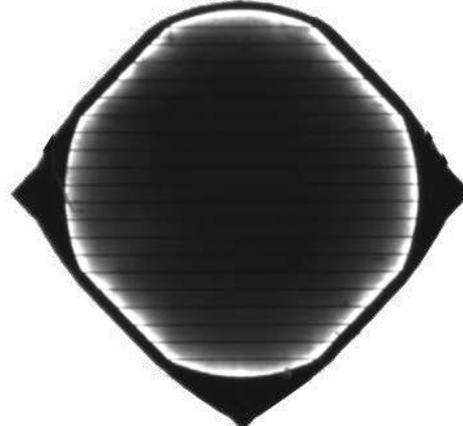 |
| Simulation (parameters of Table 3) | 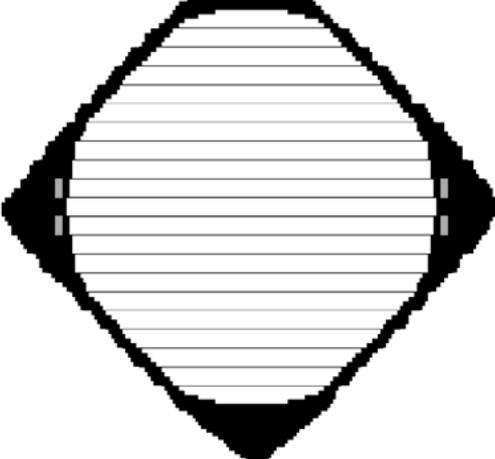 | 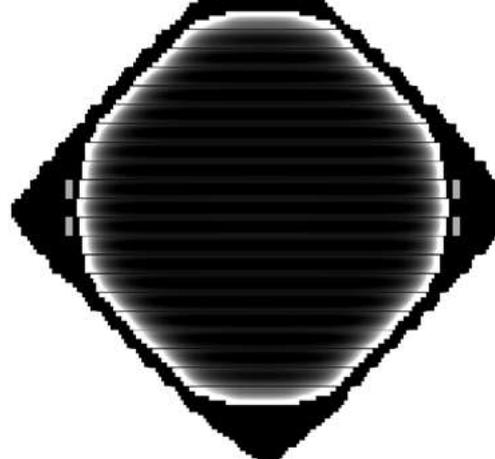 |
| Simulation (metal sheet resistance: 1 Ω/□ and specific front contact resistance: $10^{-5}$ Ω·cm$^2$) | | 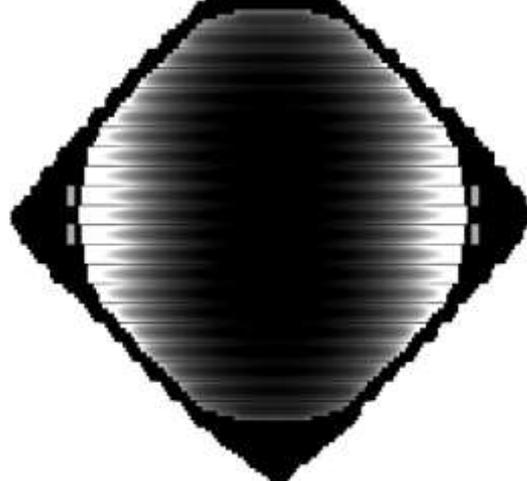 |

Figure 19. Top: Experimental EL of the GaInP top subcell. Left: "Char_Ref" solar cell. Right: "Fail High" solar cell. Center: Simulations by using the parameters of Table 3. Bottom right: Simulated EL of the GaInP top subcell of the "Fail High" solar cell using a metal sheet resistance of 1 Ω/□ and a specific front contact resistance of $10^{-5}$ Ω·cm$^2$ which fit the red curve of Figure 18. In center and bottom figures, small grey rectangles at the left and right part of the busbar emulate the wire bonding connections. Their grey color has no meaning relating EL just to highlight them from the black color of the busbar.

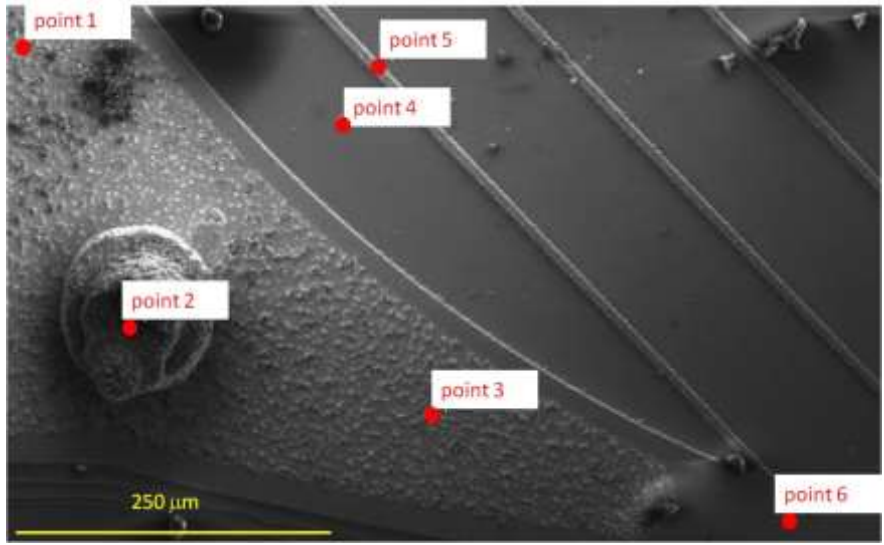

Figure 20. Planar view SEM image of the "Fail_High" solar cell surface, showing the points where EDX analysis has been performed.

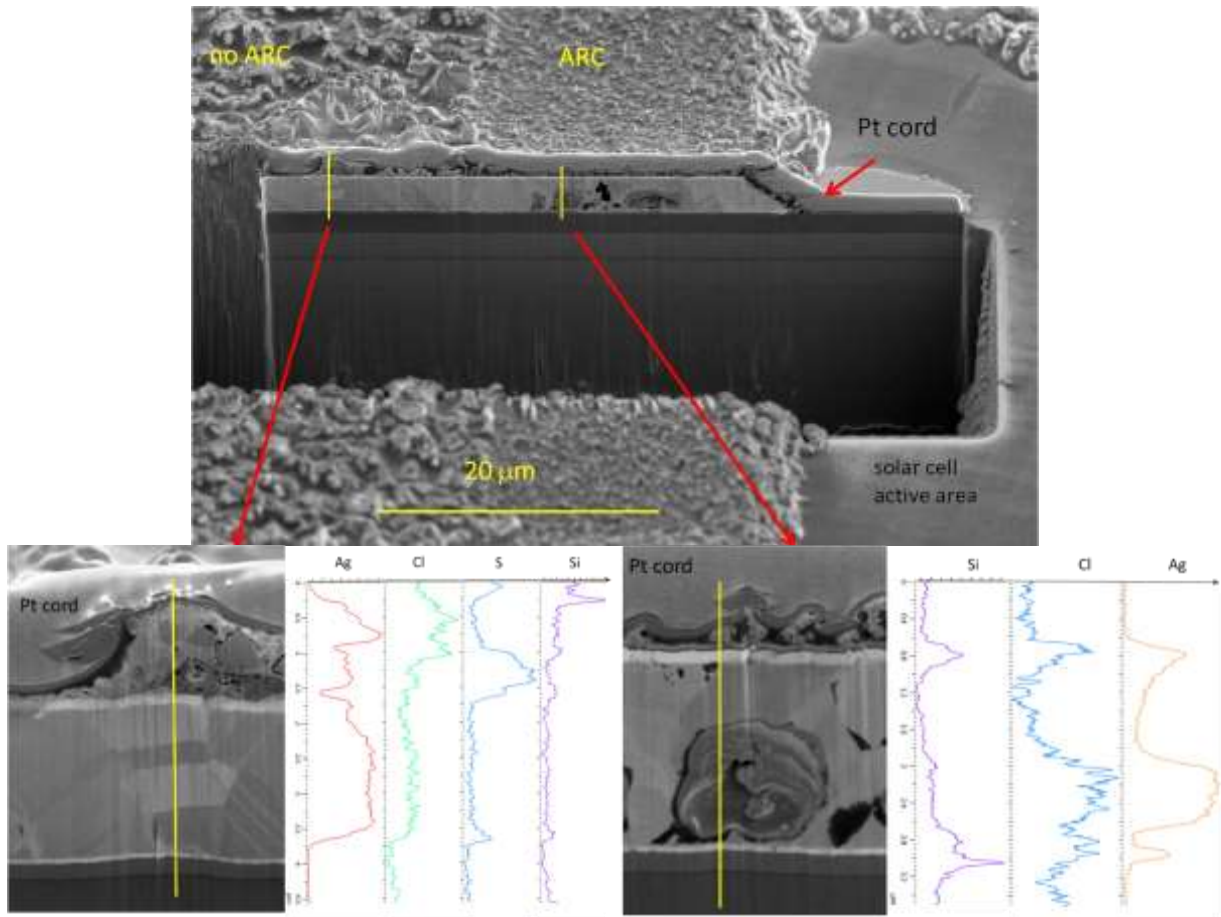

Figure 21. Top: SEM micrograph after FIB milling of the "Fail_High" solar cell busbar metallization cross-section. Pt cord is deposited in order to protect the sample surface during FIB. Bottom: (left) SEM micrograph and line EDX analysis of the busbar cross-section in the area without ARC; (right) SEM micrograph and line EDX analysis of the busbar cross-section in the area covered with ARC. Profiles of the main elements along the analyzed length (notice that there are different scale spans for each element). Yellow vertical lines correspond to the length scale on the element profiles.

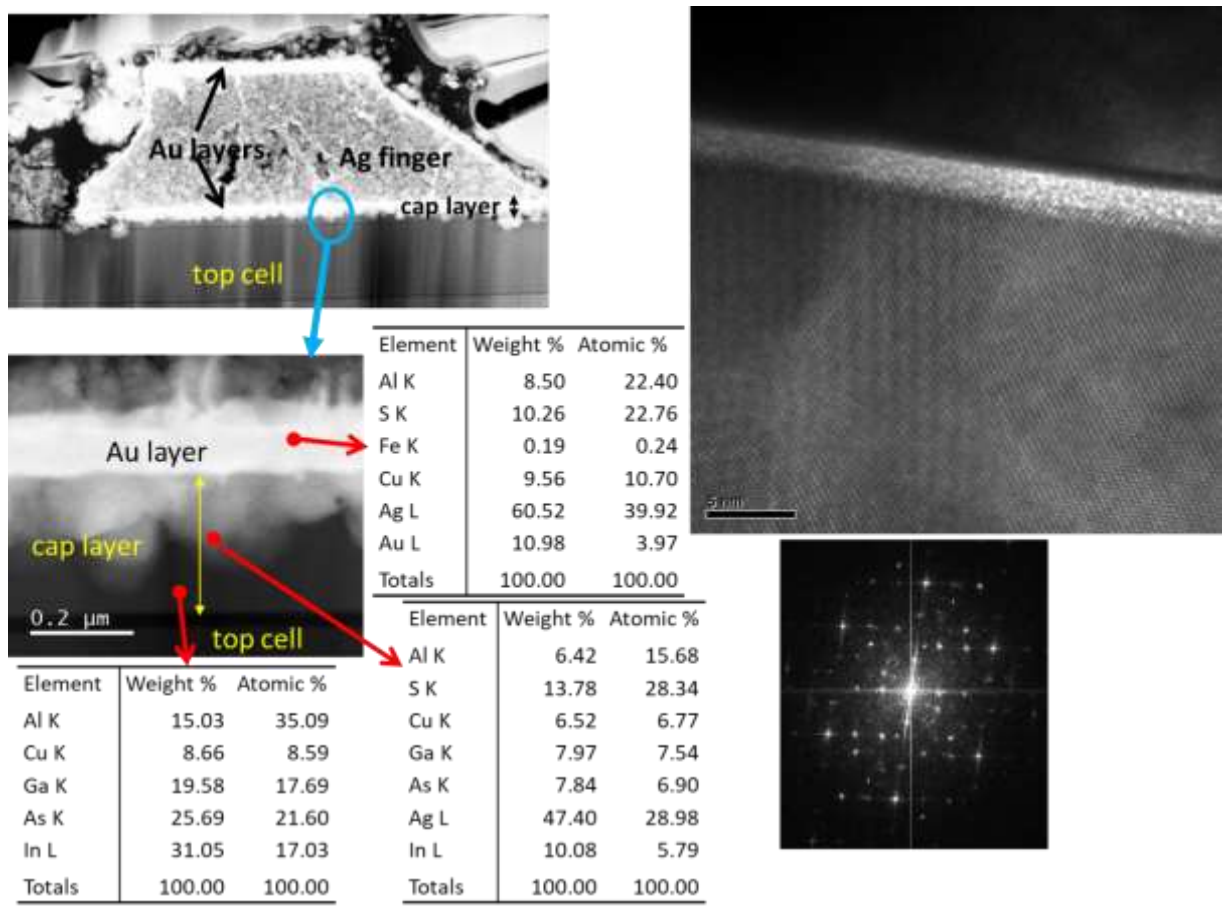

Figure 22. Cross-section view STEM image of a finger of the "Fail_High" solar cell (top left), showing the area (cyan circle) where EDX analysis has been performed (bottom left). Copper and iron presence in the EDX analysis derive from the sample holder (left). HRTEM of the Au/Ga(In)As cap layer interface (top right). Fast Fourier Transform of Ga(In)As the cap layer (bottom right).

Table 1. Nomenclature used for the solar cells analyzed in this paper. Each nomenclature presented in the table may represent several solar cell samples.

| Solar cell nomenclature | Climatic chamber temp. (°C) | $T_{cell}$ (°C) | $t_{\_Cam}$ (Hours) | $t_{\_deg}$ (Hours) | Description |
|---|---|---|---|---|---|
| Char_Ref | N/A | 25 | 0 | 0 | Reference solar cell that is used for pre- and post-characterization purposes, carefully stored and handled. |
| Ref_Low | 120 | 120 | 8900 | 0 | Reference solar cell in the 125 °C test. |
| Ref_Mid | 140 | 140 | 6207 | 0 | Reference solar cell in the 145 °C test. |
| Ref_High | 160 | 160 | 4200 | 0 | Reference solar cell in the 165 °C test. |
| Ref_IV_Low | 120 | 120 | 8900 | 0 | Reference solar cell in the 125 °C test with dark I-V Curve measurement on each degradation cycle |
| Ref_IV_Mid | 140 | 140 | 6207 | 0 | Reference solar cell in the 145 °C test with dark I-V Curve measurement on each degradation cycle |
| Ref_IV_High | 160 | 160 | 4200 | 0 | Reference solar cell in the 165 °C test with dark I-V Curve measurement on each degradation cycle. |
| Deg_Low_Rec | 120 | 125 | 6520 | 6520 | Degraded solar cell showing an increase in recombination currents, and no series resistance increase Tested at 125 °C. |
| Deg_Mid_Rs | 140 | 145 | 6207 | 6207 | Degraded solar cell showing a power loss smaller than 10% and increase in recombination currents and in series resistance. Tested at 145 °C. |
| Fail_Mid | 140 | 145 | 6207 | 6207 | Solar cell with an illumination power loss greater than 10% at 500× (failure), tested at 145 °C. |
| Fail_High | 160 | 165 | 4780 | 4780 | Solar cell with an illumination power loss greater than 10% at 500× (failure), tested at 165 °C. |

Table 2. Atomic % composition at the points marked in Figure 9.

| Spectrum | C | O | Al | P | Ti | Ga | Ag | In | Au | Total |
|---|---|---|---|---|---|---|---|---|---|---|
| point 1 | 15.80 | | | | | | 57.83 | | 26.37 | 100.00 |
| point 2 | 11.97 | | | | | | 56.63 | | 31.41 | 100.00 |
| point 3 | 14.46 | 0.74 | | | | | 60.75 | | 24.04 | 100.00 |
| point 4 | 2.56 | 40.06 | 7.18 | 22.97 | 3.38 | 8.07 | | 15.78 | | 100.00 |
| point 5 | 7.82 | 50.94 | 8.52 | | 4.18 | | 15.92 | | 12.63 | 100.00 |
| point 6 | 9.24 | 49.11 | 8.32 | | 4.54 | | 4.66 | | 24.13 | 100.00 |

Table 3. Fitting parameters for both "Char Ref" and "Fail High" cells used in the dark I-V curves of Figure 18 and EL of Figure 19. $J_{01T}$, $J_{01M}$, $J_{01B}$, $J_{02T}$, $J_{02M}$ of the "Fail High" cell have been increased 1.5 times their value for the "Char Ref" cell to fit the experimental curve. Subindexes $T$, $M$ and $B$ of $J_{01}$ and $J_{02}$ mean Top, Middle and Bottom subcells, respectively

| Parameter | "Char Ref" | "Fail High" |
|---|---|---|
| $J_{01T}$ [A/cm$^2$] | 1·10$^{-24}$ | 1.5·10$^{-24}$ |
| $J_{01M}$ [A/cm$^2$] | 4·10$^{-16}$ | 6·10$^{-16}$ |
| $J_{01B}$ [A/cm$^2$] | 5·10$^{-6}$ | 7.5·10$^{-6}$ |
| $J_{02T}$ [A/cm$^2$] | 4·10$^{-13}$ | 6·10$^{-13}$ |
| $J_{02M}$ [A/cm$^2$] | 1.4·10$^{-8}$ | 2.1·10$^{-8}$ |
| $J_{02B}$ [A/cm$^2$] | 1·10$^{-30}$ | 1·10$^{-30}$ |
| Shunt resistance [Ω·cm$^2$] | 1·10$^{6}$ | 1.4·10$^{2}$ |
| Metal sheet resistance [Ω/□] | 2.65·10$^{-2}$ | 5.3 10$^{-2}$ |
| Specific contact resistance [Ω·cm$^2$] | 1·10$^{-5}$ | 3.5·10$^{-3}$ |

Table 4. Atomic % composition at the points marked in Figure 20.

| Spectrum | C | O | Al | Si | P | S | Cl | Ti | Ga | Ag | In | Au | Total |
|---|---|---|---|---|---|---|---|---|---|---|---|---|---|
| point 1 | 18.4 | 20.0 | 1.1 | 16.8 | | 5.4 | 3.4 | | | 22.8 | | 7.5 | 100.0 |
| point 2 | 40.4 | 28.9 | | 10.7 | | | 5.2 | | | 14.9 | | | 100.0 |
| point 3 | 35.0 | 17.0 | | 5.5 | | 5.2 | 0.9 | | | 20.2 | | 16.2 | 100.0 |
| point 4 | 9.6 | 40.6 | 7.3 | 1.2 | 18.9 | | | 3.1 | 6.7 | | 12.6 | | 100.0 |
| point 5 | 26.9 | 40.9 | 6.2 | 5.0 | | | 0.9 | 3.2 | | 4.3 | | 12.6 | 100.0 |
| point 6 | 21.8 | 40.2 | 7.4 | 3.6 | | | 2.1 | 3.8 | | 6.0 | | 15.1 | 100.0 |